\documentclass[final,5p,times,twocolumn]{elsarticle}

\def \pics {./pics}      

\usepackage{graphicx}      
\usepackage{natbib}        

\usepackage{balance}
\usepackage{amsmath,amssymb,amsfonts,bm}
\usepackage{enumitem}
\usepackage{xspace}
\usepackage{url}
\usepackage{tabularx}
\usepackage{subfig}
\usepackage[algo2e, linesnumbered, algoruled]{algorithm2e}

\DeclareMathOperator{\diag}{diag}

\newcommand{\eor}{\ensuremath{\hfill\blacklozenge}}

\newcommand{\R}{{\mathbb R}}

\newcommand{\N}{{\mathbb N}}
\newcommand{\I}{{\cal I}}
\newcommand{\be}{\begin{equation}}
\newcommand{\ee}{\end{equation}}
\newcommand{\ba}{\begin{array}}
\newcommand{\ea}{\end{array}}
\newcommand{\baa}{\left[\begin{array}}
\newcommand{\eaa}{\end{array}\right]}

\newcommand{\beqa}{\begin{eqnarray}}
\newcommand{\eeqa}{\end{eqnarray}}
\newcommand{\bt}{\begin{tabular}}
\newcommand{\et}{\end{tabular}}

\newcommand{\bi}{\begin{itemize}}
\newcommand{\ei}{\end{itemize}}
\newcommand{\ben}{\begin{enumerate}}
\newcommand{\een}{\end{enumerate}}
\newcommand{\bc}{\begin{center}}
\newcommand{\ec}{\end{center}}

\newcommand{\bse}{\begin{subequations}}
\newcommand{\ese}{\end{subequations}}

\newtheorem{rem}{Remark}
\newtheorem{assum}{Assumption}

\newcommand{\figref}[1]{Fig.~\ref{#1}\xspace}

\newcommand{\secref}[1]{Section~\ref{#1}\xspace}

\newcommand{\remref}[1]{Remark~\ref{#1}\xspace}

\newcommand{\algref}[1]{Algorithm~\ref{#1}\xspace}

\renewcommand{\algref}[1]{Algorithm~\ref{#1}\xspace}%

\newcommand{\norm}[1]{\left\lVert#1\right\rVert}
\newcommand{\abs}[1]{\left|#1\right|}
\newcommand{\eref}[1]{(\ref{#1})}

\begin{document}
\begin{frontmatter}

\title{Fault Handling in Large Water Networks with Online Dictionary Learning}

\author[1]{Paul Irofti\corref{cor1}\fnref{fn1}} 
\ead{paul@irofti.net}

\author[2]{Florin Stoican}
\ead{florin.stoican@acse.pub.ro}

\author[3]{Vicenç Puig}
\ead{vicenc.puig@upc.edu}

\address[1]{Department of Computer Science and the Research Institute of the University of Bucharest (ICUB), Romania.}

\address[2]{Department of Automatic Control and Computers, University Politehnica of Bucharest, Romania}

\address[3]{Universitat Politècnica de Catalunya, Institut de Robòtica i Informàtica Industrial (CSIC, UPC), Barcelona, Spain}

\cortext[cor1]{Corresponding author}
\fntext[fn1]{This work was supported by the Romanian National Authority for Scientic Research, CNCS - UEFISCDI, project number PN-III-P1-1.1-PD-2019-0825.}

\begin{abstract}                

Fault detection and isolation in water distribution networks is an active topic due to the nonlinearities of flow propagation and recent increases in data availability due to sensor deployment.
Here, we propose an efficient two-step data driven alternative:
first,
we perform sensor placement taking the network topology into account; second,
we use incoming sensor data
to build a network model through online dictionary learning.
Online learning is fast and
allows tackling large networks as it processes small batches of signals at a time.
This brings the benefit of
continuous integration of new data into the existing network model,
either in the beginning for training
or in production when new data samples are gathered.
The proposed algorithms show good performance in our simulations on both small and large-scale networks.


\end{abstract}


\end{frontmatter}

\section{Introduction}
Water distribution networks (along electricity, transport and communication ones) are a critical infrastructure component. Thus, modeling and observability issues are of paramount importance and have to be handled through increases in connectivity, automation and smart metering. 

In particular, pipe leakages (assimilated hereinafter with fault events) have to be detected and isolated as soon and as precisely as possible. While apparently straightforward (the leakage leads to a measurable loss of pressure), several issues conspire in increasing the scheme's difficulty:
\begin{enumerate}[label=\roman*),nosep]
    \item disparity between the number of nodes (hundreds /thousands in a large-scale network) and of sensors (expensive/hard to install, hence at most tens) \cite{Zhao18_dynamic, perelman2016sensor,xu2008identifying};
    \item water network dynamics are strongly nonlinear and demand uncertainties are significant \cite{sela2018robust,brdys1996operational}.
\end{enumerate}
Hence, sensors have to be placed to provide network-wide relevant information while pressure and flow information is obtained either through emulation \cite{epanet} or experimentally \cite{perez2014leak}. Such data driven analysis naturally leads to heuristic implementations which come with specific caveats:
\begin{enumerate}[label=\roman*),nosep]
    \item heuristic methods use the data agnostically and ignore information about network structure/particularities;
    \item network size may lead to implementations which clog the resources or are bogged by numerical artifacts.
\end{enumerate}




In light of the previous remarks, it is clear that the main issues are sensor placement and subsequent fault detection and isolation (FDI) mechanism. For the former, we propose a novel Gram-Schmidt graph-aware procedure and for the later we consider a dictionary learning (DL) approach.

Specifically, we assign the faults affecting a given node to a class and train a dictionary such that its atoms discriminate between these classes. The subsequent classification of a signal in terms of the dictionary's atoms serves as proxy for FDI. The active atoms for a certain class are seen as a fault signature which unambiguously asserts FDI (if the signature is unique w.r.t. the other possible faults).

DL~\cite{DL_book}
is an active research topic in the signal processing community
providing machine learning algorithms
that build linear models based on the given (obtained from processes with nonlinear dynamics) input data.
Its applications to classification tasks~\cite{JLD13} in general
and online classification~\cite{IB19_toddler} in particular
provide fast and memory efficient implementations 
well suited for IoT devices and for online production usage.
Furthermore, DL methods are known to be robust to various types of noises and perturbations~\cite{JGF12_noiseDL}, a key property for our application.

Our previous work~\cite{IS17_ifac, SI19_mpdl} has shown encouraging results when adapting DL classification for FDI in water networks.
In this paper, we propose new methods that tackle large distribution networks and lift data dimensionality limitations by employing online DL strategies (which process data in small batches which translate into smaller computation complexities).
This leads to a small decrease of the FDI performance as compared to a method handling the whole data set at once, however
this is quickly attenuated as more data is processed.




In simulations, we consider both the proof-of-concept benchmark ``Hanoi network'' and a generic large-scale network~\cite{muranho2012waternetgen}. Furthermore, we use multiple demand profiles, fault magnitudes and discuss different sensor placement strategies and success criteria. 


\section{Preliminaries}
\label{sec:prelim}
A passive water network (i.e., without active elements like pumps) consists of one or more tank nodes (whose heads\footnote{In the water network parlance, ``head'' denotes the height of the column of water in a node wrt a common ground level.} remain constant) which feed a collection of junction nodes through a network of interconnected pipes. From a modeling perspective, the question is what are the flows through the pipes, what are the heads through the junction nodes and how do these variables depend on user demand (outflows from some or all of the junction nodes) and unexpected events (in our case: pipe leakages). 

\subsection{Steady-state behavior}

The dynamics of the network are usually ignored. This is a reasonable assumption as long as demand variation is slow and unexpected events (e.g., leakages) are rare. In other words, any transient-inducing event is sufficiently rare and the transients themselves are sufficiently fast such that it is a fair approximation to consider the system at equilibrium \cite{brdys1996operational}. Since water is incompressible  the relevant physical laws which apply are those of mass and energy conservation. First, the inflows and outflows passing throughout a junction node have to balance:
\be
\label{eq:incidenceflow}
\sum\limits_{j=1}^{n} \mathbf B_{ij}\mathbf q_j=\mathbf c_i
\ee
where $\mathbf q_j$ is the flow through pipe $j$, $\mathbf c_i$ is the consumption of node $i$ and $\mathbf B$ is the adjacency matrix of the network, i.e., $\mathbf B_{ij}$ takes one of the following values:
\be
\label{eq:b}
\mathbf B_{ij}=\begin{cases}\hphantom{-}1,& \textrm{if pipe $j$ enters node $i$;}\\ \hphantom{-}0,& \textrm{if pipe $j$ is not connected to node $i$;}\\ -1,&\textrm{if pipe $j$ leaves node $i$.}\end{cases}
\ee
Next, the empiric Hazen-Williams formula \cite{sanz2016demand} gives the head flow variation between nodes $i,j$ linked through a pipe with index $\ell$ (we assume that the pipe of index $\ell$ links the $i$, $j$-th nodes):
\be
\label{eq:hw}
\mathbf h_i-\mathbf h_j=\frac{10.67\cdot L_\ell}{C_\ell^{1.852}\cdot D_\ell^{4.87}}\cdot \mathbf q_\ell\cdot |\mathbf q_\ell|^{0.852}
\ee
where $L_\ell$ is the length in $[m]$, $D_\ell$ is the diameter in $[m]$ and $C_\ell$ is the adimensional pipe roughness coefficient; the flow $\mathbf q_\ell$ is measured in $[m^3/s]$. 

Using \eqref{eq:hw} we express the flow in terms of the associated head flow $\mathbf h_i-\mathbf h_j$:
\be
\label{eq:headflow}
\mathbf q_{\ell}=G_{ij}^{0.54}(\mathbf h_i-\mathbf h_j)|\mathbf h_i-\mathbf h_j|^{-0.46} 
\ee
where $G_{ij}$ is the pipe conductivity, defined as
\be
G_{ij}=\frac{1}{R_{ij}}=\frac{C_\ell^{1.852}\cdot D_\ell^{4.87}}{10.67\cdot L_\ell}.
\ee 
Noting that the $\ell$-th line of the column vector $-\mathbf B^\top \mathbf h$ returns the difference $\mathbf h_i-\mathbf h_j$ and combining \eqref{eq:incidenceflow} with \eqref{eq:headflow} leads to the nonlinear steady-state relations:
\be
\label{eq:steadystate}
\mathbf B\mathbf G\left[\left(-\mathbf B^\top \mathbf h+\mathbf B_f^\top \mathbf h_f\right)\times \left|-\mathbf B^\top \mathbf h+\mathbf B_f^\top \mathbf h_f\right|^{-0.46}\right]=\mathbf c,
\ee
where $\mathbf G=\diag \left(G_\ell\right)$ and `$\times$' denotes the elementwise multiplication of two vectors (i.e., the i-th element of $x\times y$ is $[x\times y]_i=x_iy_i$). Note the addition of term $\mathbf B_f^\top \mathbf h_f$ which describes the influence of fixed-head nodes (the tanks which feed the network). For further use, we denote with $N$ the number of junction nodes.

\subsection{Node consumption}

Assuming that all the parameters of \eqref{eq:steadystate} are known (gathered in the left side of the equation), there still remains the node consumption $\mathbf c$ as a source of uncertainty. Historically user demand data has been acquired sparsely or not at all. Most common approaches are to consider a consumption profile (usually with a week-long period) and scale it wrt the total consumption in the network:
\be
\label{eq:dnominal}
\mathbf c_i(t)=\frac{\bar {\mathbf c}_i}{\sum\limits_j \bar {\mathbf c}_j}\cdot p(t)\cdot \mathbf q_{in}(t)+\mathbf \eta_i(t),
\ee
where $p(t)$ and $\mathbf{q}_{in}(t)$ are the consumption profile and, respectively, the total water fed to the network at time instant $t$; $\bar {\mathbf c}_i$ denotes the base demand for the i-th node and $\mathbf \eta_i(t)$ covers `nominal' (those occurring under healthy functioning) uncertainties affecting the i-th node (without being exhaustive: normal user variations, seasonal and holiday variations, small leakages). 

The issue of interest is how to detect and isolate a pipe leakage. We note that a pipe leakage means a loss of flow and thus a loss of head in the network's nodes. We then interpret pipe leakage as an additive term in the node consumption value\footnote{Hereafter when we speak about isolating a leakage we refer to identifying the node directly affected by the pipe leakage. The actual leakage isolation means checking the pipes which enter into the node.}:
\be
\label{eq:dfaulty}
\mathbf c_i(t)=\frac{\bar {\mathbf c}_i}{\sum\limits_j \bar {\mathbf c}_j}\cdot p(t)\cdot \mathbf q_{in}(t)+\mathbf \eta_i + \mathbf f_i.
\ee
For further use, we consider that the profile $p(t)$ can take values from a collection of $P$ profiles $\{p_1(t),\dots, p_P(t)\}$. 
\begin{rem}
\label{rem:steady}
This means that the active profile in \eqref{eq:dnominal}--\eqref{eq:dfaulty} may be unknown at measuring time. This additional uncertainty may hide water losses due to pipe leakages. The preferred solution is in practice to measure total water demand $\mathbf q_{in}(t)$ at times when user demand is minimal (middle of the night). At this time, deviations due to leakages represent a larger percentage from the total consumption (wrt the uncertainty due to the profile). Thus, a change from the expected value may signify that leakages are present (in FDI parlance, a fault is detected). \eor
\end{rem}

\begin{figure*}[ht!]
\centering
\includegraphics[width=1.75\columnwidth]{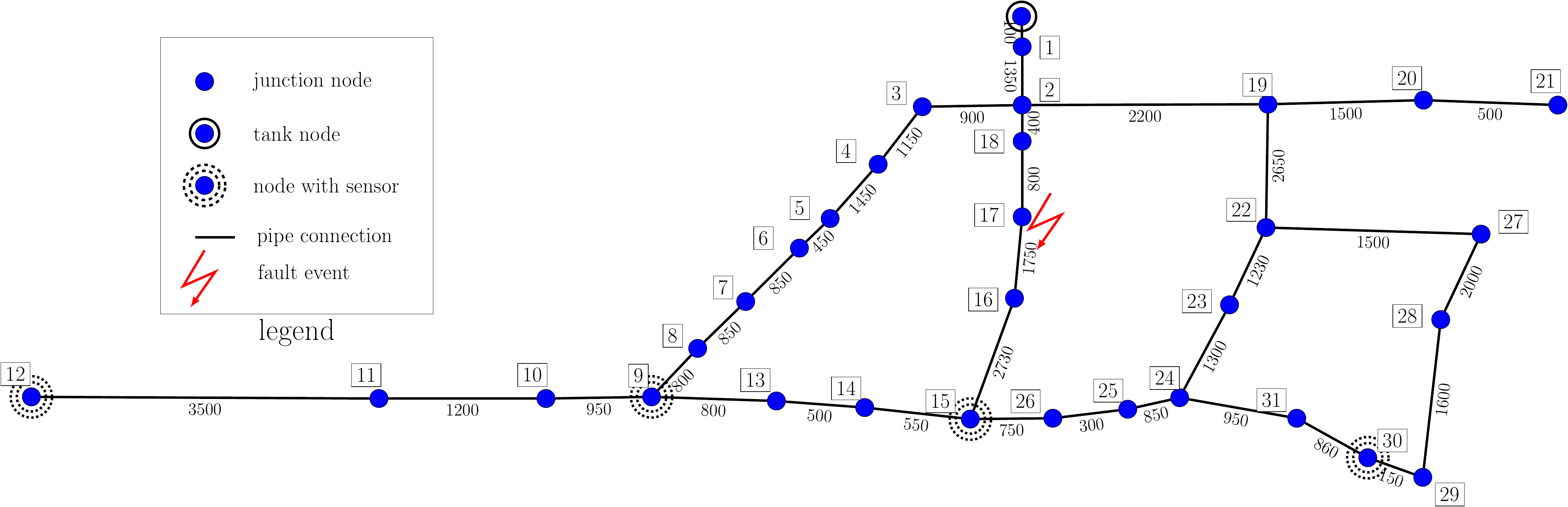}
\caption{Hanoi water network.}
\label{fig:hanoi}
\end{figure*}
\subsection{Leakage isolation and residual generation}

The issue of leakage isolation still remains. To asses the leakage events we have to compare the ``healthy'' (nominal) behavior, as given in \eqref{eq:dnominal}, with the measured (and possibly faulty, as given in \eqref{eq:dfaulty}) behavior of the network's nodes. This is done through a \emph{residual signal} which is \cite{blanke2006diagnosis}: i) constructed from known quantities; ii) sensitive to a fault occurrence\footnote{Hereinafter, to keep with the FDI context we denote a `leakage event' as a `fault occurrence'.}; and iii) robust, in as much as is possible, to normal variations. 

For further use we make a couple of assumptions.
\begin{assum}
We consider that there are no multiple fault occurrences in the network (i.e., the network is either under nominal functioning or with a single node under fault). \eor
\end{assum}
\begin{assum}
Without loss of generality we assume that the fault magnitude values are the same for each node and are taken from a finite collection ($M$  possible values from $\{m_1,\dots m_M\}$). \eor
\end{assum}

For further use we consider the nodes' head as a proxy for fault occurrences and use its nominal ($\bar{\mathbf h}$) and measured values ($\hat{\mathbf h}$) to construct the residual signal. The following aspects are relevant:
\begin{itemize}
\item as per \remref{rem:steady}, we consider an interval $K$ in which the head values remain relatively constant and average over it to obtain the ``steady-state'' nominal / measured head values:
\be
\label{eq:average_head}
\bar{\mathbf h}=\frac{1}{|K|}\sum\limits_{k\in K} \bar{\mathbf h}[k], \quad \hat{\mathbf h}=\frac{1}{|K|}\sum\limits_{k\in K} \hat{\mathbf h}[k].
\ee
\item the residual may be defined in absolute or relative form, i.e.: 
\be
\label{eq:residual}
\mathbf r^A=\mathbf{\hat h} - \mathbf{\bar h}, \quad \mathbf r^R=\frac{\mathbf{\hat h} - \mathbf{\bar h}}{\mathbf{\bar h}}.
\ee
Whenever the residuals' type (absolute or relative) are not relevant we ignore the superscripts $`A,R'$.
\item assuming that the i-th node is under fault with magnitude $m_j$ and that the network functions under profile $\mathbf p_k$, we note\footnote{With this notation, the nominal head, $\bar{\mathbf h}$, would be denoted as $\bar{\mathbf h}^{\bar k}$, where $p_{\bar k}$ is the profile active when the nominal head was measured. Since the nominal head remains constant (we cannot assume that $\bar k$ is updated), we keep the simpler notation $\bar{\mathbf h}$.} the head values with $\hat{\mathbf h}_{ij}^k$ and corresponding residual with $\mathbf r_{ij}^k$.
\end{itemize}

For further use we gather all the residual vectors $\mathbf r_{ij}^k$ into the \emph{residual matrix}\footnote{Taking all possible combinations, the residual matrix has $D=N\cdot M\cdot P$ columns. For large-scale networks or if arbitrary selections of profiles, faults and fault magnitudes are considered, the value of $D$, and consequently, the arranging and content of $\mathbf R$, may differ.} $\mathbf R\in \mathbb R^{N\,\times\, D}$:
\begin{multline}
\label{eq:resmat}
\mathbf R= \bigg[\underbrace{\mathbf r_{11}^1\,\mathbf r_{12}^1\,\dots\,\mathbf r_{1M}^1\, \dots \,\mathbf r_{11}^P\, \mathbf r_{12}^P\,\dots \mathbf r_{1M}^P}_{f_1}\dots \\ \underbrace{\mathbf r_{N1}^1\, \mathbf r_{N2}^1\,\dots \mathbf r_{NM}^1\, \dots \,\mathbf r_{N1}^P\, \mathbf r_{N2}^P\,\dots \mathbf r_{NM}^P}_{f_N}\bigg]
\end{multline}
\begin{rem}
The residual ordering inside the matrix is not essential. It was chosen thus to make easier the grouping after fault occurrence (all cases which correspond to a certain node under fault are stacked consecutively).\eor
\end{rem}


\subsection*{The Hanoi benchmark}
To illustrate the aforementioned notions, we consider a often-used benchmark in the literature: the Hanoi water network \cite{casillas2013optimal}. As seen in \figref{fig:hanoi}, the network characteristics are: one tank and $31$ junction nodes linked through $34$ pipes (each with its own length and diameter); each junction node can be affected by a leakage and some of the nodes will have sensors mounted on them. 

\begin{figure}[ht!]
\centering
\subfloat[profile]{\label{fig:hanoi_transitional_a}\includegraphics[width=.85\columnwidth]{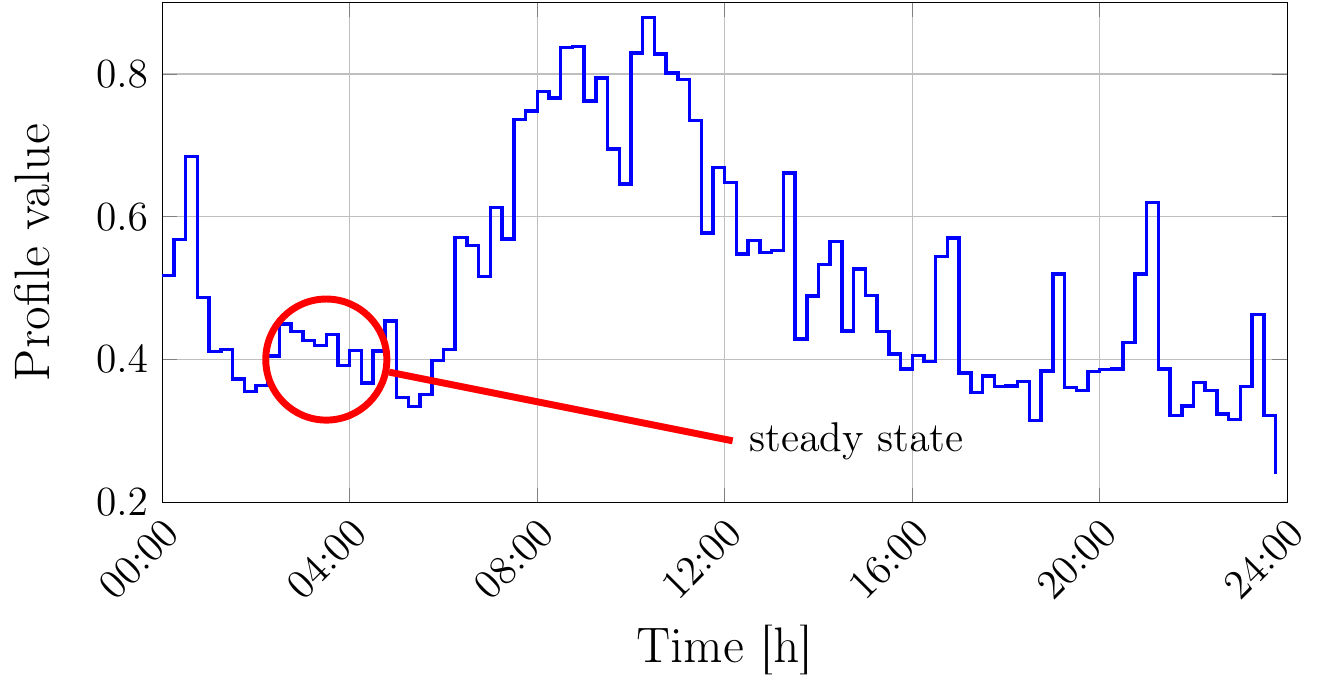}}\\
\subfloat[dynamical head response]{\label{fig:hanoi_transitional_b}\includegraphics[width=.85\columnwidth]{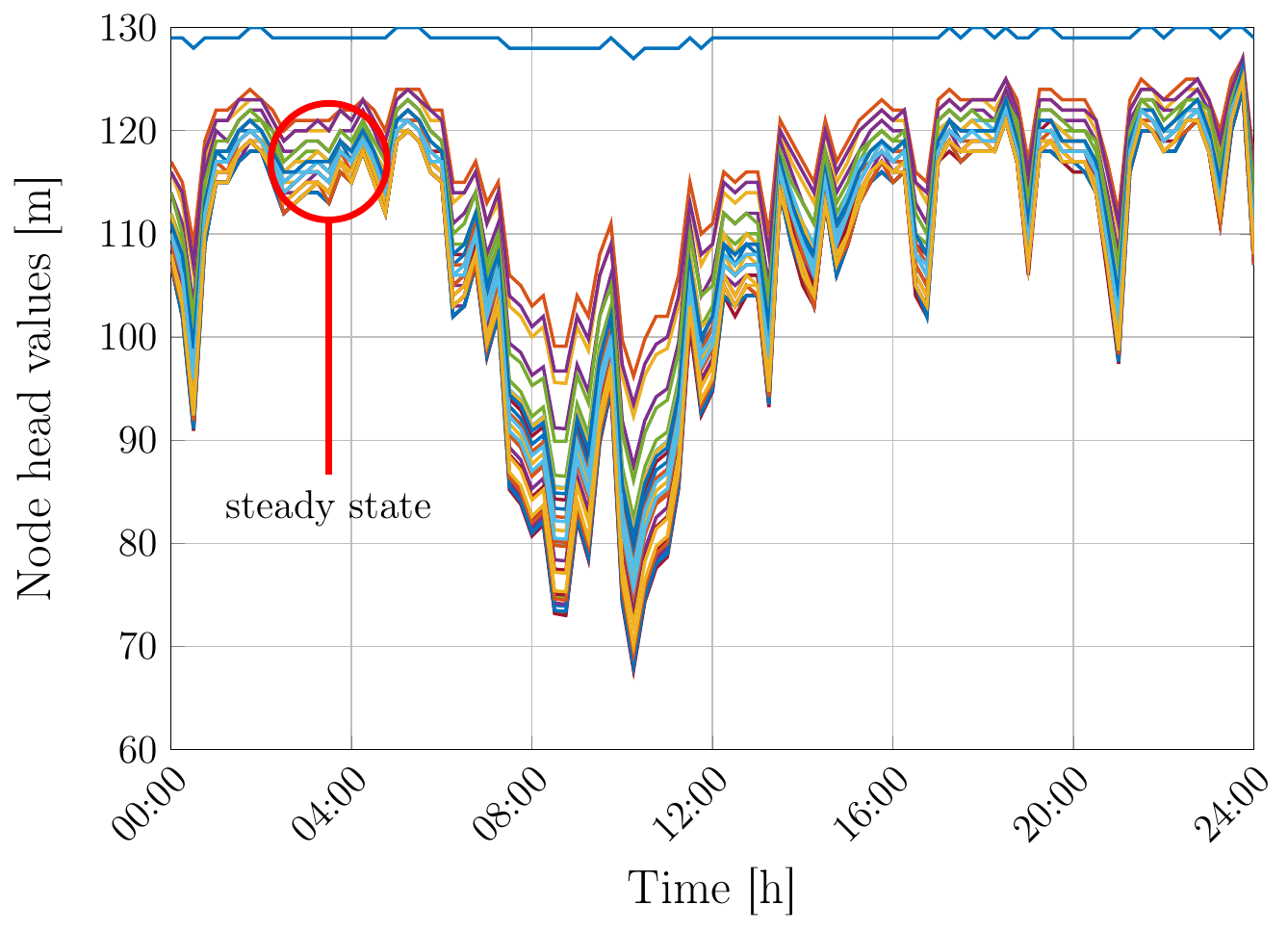}}
\caption{The head response for a given profile in the Hanoi network.}
\label{fig:hanoi_transitional}
\end{figure}

With the network profile (which multiplies each of the junction nodes' base demand) given in \figref{fig:hanoi_transitional_a}, we simulate the nodes' head under nominal functioning for one day (with 15 minutes sampling) through the EPANET software \cite{epanet}, as seen in \figref{fig:hanoi_transitional_b}. 
We observe that the empiric rule from \remref{rem:steady} holds: the head values remain steady around 3 AM, thus justifying the choice of constructing the head values with information from this time interval.

Further, we consider 9 additional perturbations (hence $P=10$) of the nominal profile shown in \figref{fig:hanoi_transitional_a} through the addition of uniform noise bounded in the range of $\pm 2.5\%$. To illustrate the fault effects we consider such an event at node $17$ with fault magnitudes taken from $\{0,4,8,12,20\}$, hence $M=5$. Furthermore, for each fault magnitude we run $P=10$ times the EPANET software (once for each flow profile).
\begin{figure}[ht!]
\centering
\label{fig:hanoi_residual_a}\includegraphics[width=.85\columnwidth]{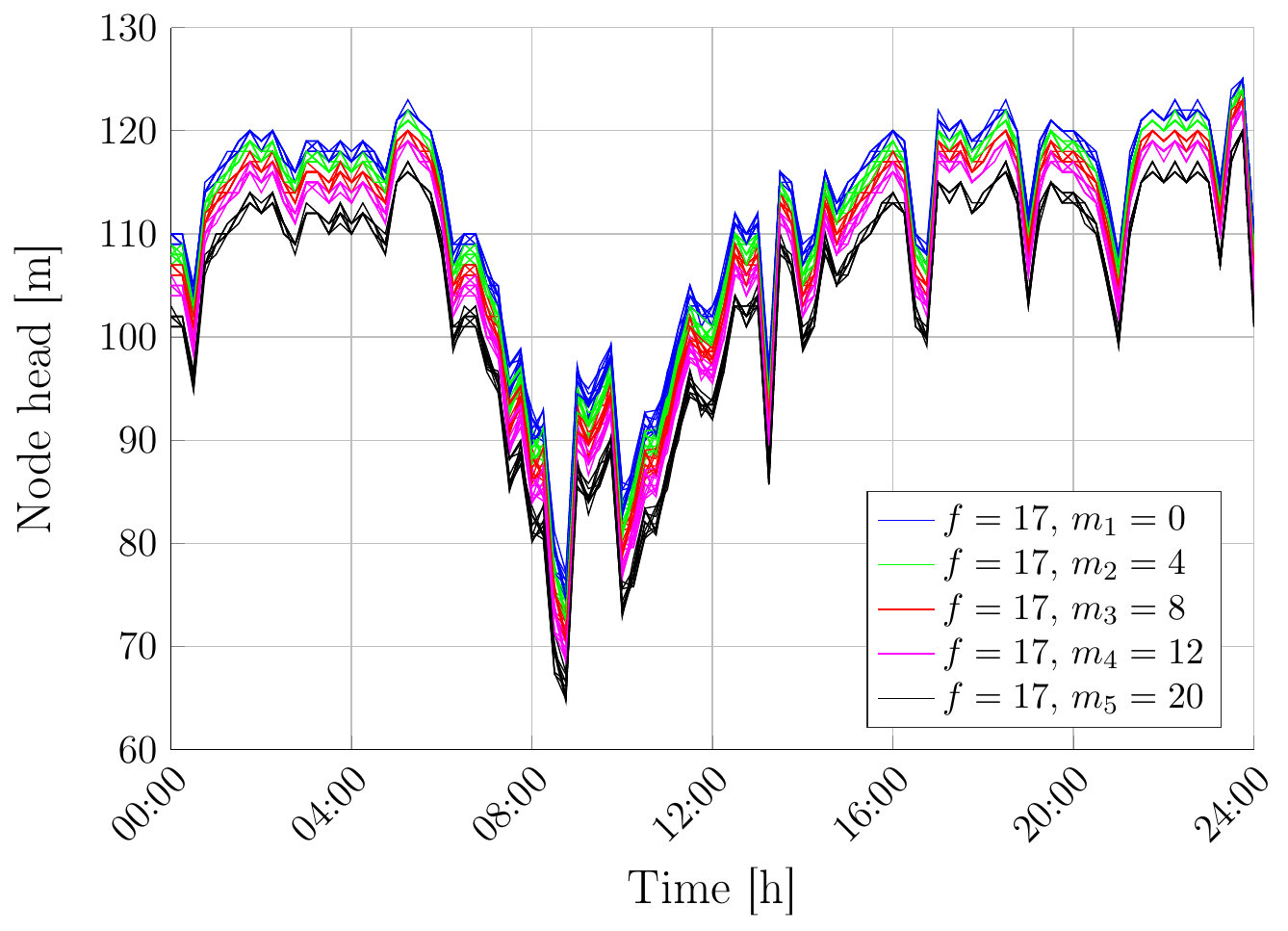}
\caption{Illustration of junction node $17$ head variation while under fault and with various flow profiles and fault magnitudes.}
\label{fig:node_under_fault}
\end{figure}
The resulting head values are shown (each group of plots with the same color denotes the node's value under the same fault magnitude but for different profiles) in \figref{fig:node_under_fault} where we can observe, as expected, that the fault affects the node's head value.

\begin{figure}[ht!]
\centering
\includegraphics[width=.85\columnwidth]{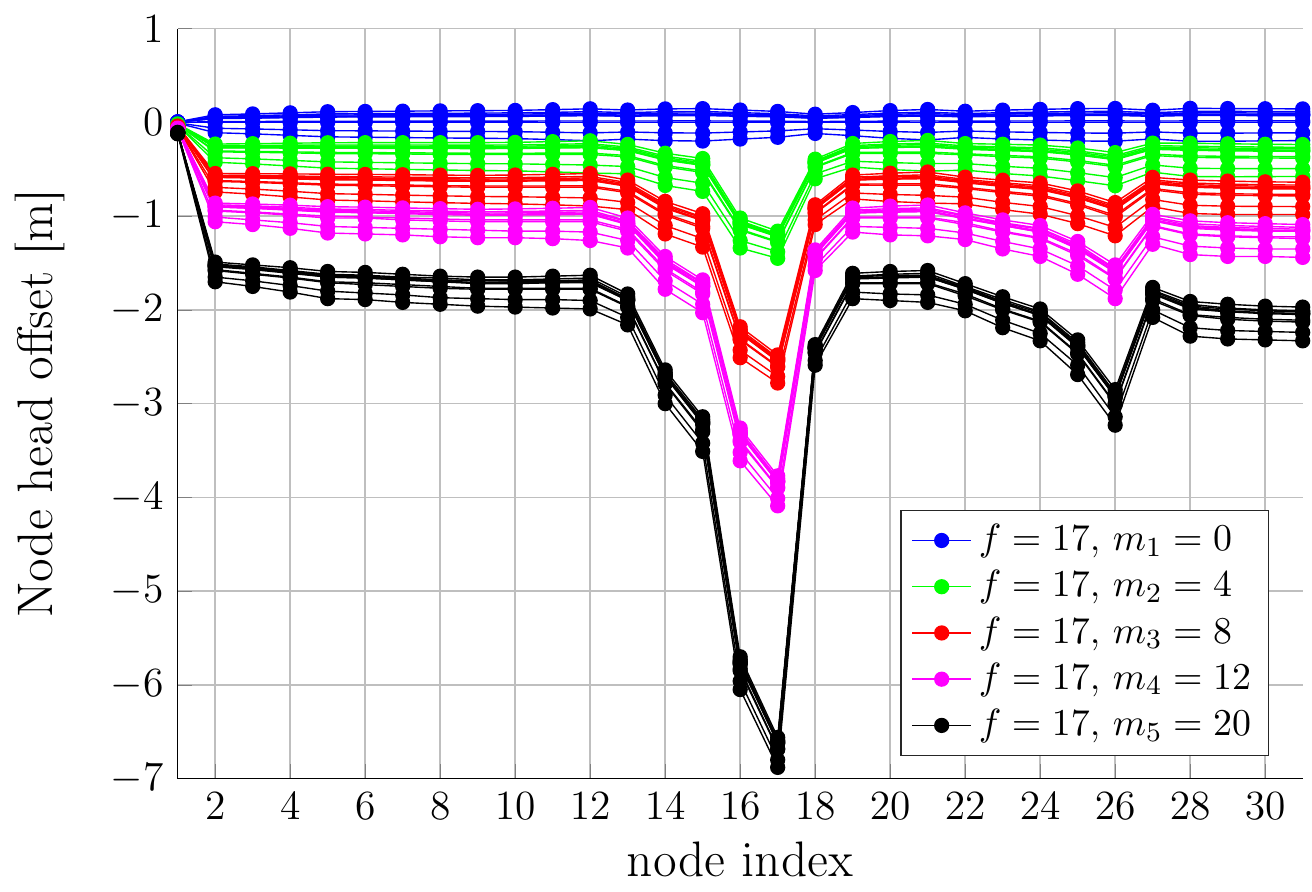}
\caption{Residual vectors in the Hanoi network.}
\label{fig:hanoi_residual}
\end{figure}

Taking $K=[12,18]$ and using it as in \eqref{eq:average_head} to obtain the residuals \eqref{eq:residual} leads to the plots shown in \figref{fig:hanoi_residual} (we consider the absolute residual variant).

As expected, node 17 (where the fault happens) is the most affected. Still, measurable effects appear in nodes 14, 15 or 26. This is noteworthy for the subsequent fault detection/isolation analysis as it shows fault propagation throughout the network. 

\subsection{Problem Statement}

The main idea is to detect and isolate fault occurrences (i.e., leakages) within the water network with a limited amount of information (much fewer sensors than nodes). Due to its complexity (network size, nonlinearities, demand uncertainty, etc.), the problem is divided into two consecutive steps:
\begin{enumerate}[label=\roman*),nosep]
\item the sensor placement, i.e., where to place the limited number of sensors such that the subsequent fault detection and isolation is maximized; 
\item the  fault detection and isolation procedure which provides an estimation of the fault occurrences (their location within the network).
\end{enumerate}
The ideas, building on \cite{IS17_ifac}, are to provide a dictionary learning framework within which to:
\begin{enumerate}[label=\roman*),nosep]
\item implement a Gram-Schmidt procedure which uses the gathered data to propose candidate nodes for sensor placement;
\item onto the reduced residual data, apply an online dictionary learning procedure which first trains a dictionary (an overcomplete basis for the residual signals) which is further used to classify test residuals into one of the predefined classes. Associating each class to a fault event means that the classification step itself becomes the fault detection and isolation mechanism.
\end{enumerate}
Both elements exploit the network's structure and manage its numerical complexities: the network's Laplacian penalizes the sensor placement described in \secref{sec:placement} and the online DL implementation described in Section~\ref{sec:dictionary} allows to handle large datasets (otherwise cumbersome or downright impossible through other, offline, procedures).

\section{Sensor Placement}
\label{sec:placement}

Arguably, the main difficulty in maximizing the network's observability (and, thus, improve the FDI mechanism) comes from inadequate sensor placement: no subsequent FDI mechanism, (regardless of its prowess) can overcome the handicap of inadequate input data.

The problem reduces in finding a sequence of indices $\mathcal I\subset \{1\dots N\}$ with at most $s$ elements from within the list of available node indices such that the FDI mechanism provides the best results. As formulated, the problem has a two-layer structure: at the bottom, the FDI mechanism is designed for a certain sensor selection and at the top, the sensor selection is updated to reach an overall optimum. The nonlinearities and large scale of the problem mean that we have to break it into its constituent parts: first the sensors are placed (based on available data and/or model information) and, subsequently, the FDI mechanism is optimized, based on the already computed sensor placement. 

While there are multiple approaches in the literature, the sensor placement problem is still largely open. One reason is that the degree to which a node is sensitive to a leak is roughly proportional with the inverse of its distance from the leaky node. Therefore, any selection strategy which does not use the entire information provided by the network is biased towards clumping sensor locations. On the other hand, analytic solutions which consider the network as a whole are computationally expensive (or downright impractical).  

These shortcomings have motivated many works for various large-scale networks \cite{kim2018pmu,krause2008near} as well as for water networks specifically \cite{meseguer2014decision,perelman2016sensor,zan2014event}. While varied, the approaches can be grouped into \cite{sela2018robust}: i) mixed-integer or greedy procedures which solve some variation of the set cover problem; ii) evolutionary algorithms which employ heuristic methods; and iii) topology-based methods which make use of the underlying graph structure of the network.

In particular, from the class of greedy procedures we analyze two minimum cover variants (set/test covering) and from the class of topology-based methods we propose an extended Gram-Schmidt procedure with Laplacian regularization.



\subsection{Minimum cover procedures}

Let us consider the residual matrix $\mathbf R$ defined as in\footnote{In fact, we use only a subset of columns from \eqref{eq:resmat}, the so-called \emph{training set}, but for simplicity we abuse the notation.} \eqref{eq:resmat}. To this matrix corresponds the \emph{fault signature matrix} $\mathbf M$ obtained from the former by a binarization procedure \cite{blanke2006diagnosis}:
\be
\label{eq:binarization}
\mathbf M_{ij}=\begin{cases}1, & \exists k:\: \textrm{s.t. } k \textrm{ corresponds to fault }j \textrm{ and }|\mathbf R_{ik}|\geq \tau,\\ 0, &\textrm{otherwise}.\end{cases}
\ee
\eqref{eq:binarization} should be read as follows: if any\footnote{The ``any'' condition can be replaced with any selection criterion deemed necessary (e.g., ``all entries'', ``the majority of the entries'').} of the entries of the $i$-th node which correspond to the active fault $j$ are above a pre-specified threshold $\tau$ then the fault $j$ is detected by the node $i$ (i.e., $\mathbf M_{ij}=1$). 

With $\mathbf M$, the fault signature matrix, given as in \eqref{eq:binarization}, we apply the \emph{minimum set cover (MSC)} procedure from \cite{perelman2016sensor}, in a variation of the mixed-integer form appearing in \cite{sela2018robust}:
\bse
\label{eq:mi_all}
\begin{align}
\label{eq:mi_a}\arg\max\limits_{\gamma_j,\alpha_i}\quad &\sum\limits_{j\in \mathcal F} \gamma_j\\
\label{eq:mi_b}\textrm{s.t.} & \sum\limits_{i\in \mathcal N} \mathbf M_{ij}\alpha_i\geq \gamma_j, \: j \in \mathcal F\\
\label{eq:mi_c}& \sum\limits_{i \in \mathcal N} \alpha_i \leq s\\
\label{eq:mi_d}& 0\leq \gamma_j\leq 1, \: j \in \mathcal F\\
\label{eq:mi_e}& \alpha_i \in \{0,1\}, \: i \in \mathcal N.
\end{align}
\ese
$\mathcal F$, $\mathcal N$ denote (in our case, $\mathcal F=\mathcal N=\{1,\dots, N\}$) the list of faults and nodes, respectively. Parameter $s$ limits the number of available sensors. Taking $\gamma_j=1$, \eqref{eq:mi_all} reduces to finding $\alpha_i$ such that \eqref{eq:mi_b}, \eqref{eq:mi_c} and \eqref{eq:mi_e} hold: \eqref{eq:mi_b} ensures that each fault $j$ is detected by at least a node $i$; \eqref{eq:mi_c} ensures that at most $s$ selections are made and \eqref{eq:mi_e} ensures that the selection is unambiguous (a node is either selected, $\alpha_i=1$ or not, $\alpha_i=0$). This formulation may prove to be infeasible (there might be no node selection which permits complete fault detection), thus requiring the addition of the slack variables $\gamma_j$, their constraining in \eqref{eq:mi_d} and subsequent penalization in the cost \eqref{eq:mi_a}.
 
As noted in \cite{perelman2016sensor}, \eqref{eq:mi_all} maximizes fault detection but does not guarantee fault isolation (an ideal solution to \eqref{eq:mi_all} would be to find a unique node which detects all faults; this is, obviously, unfortunate from the viewpoint of fault isolation). The solution proposed in \cite{perelman2016sensor}, at the cost of greatly expanding the problem size, is to construct an auxiliary matrix $\tilde{\mathbf M}\in \{0,1\}^{|\mathcal N| \times {{|\mathcal F|}\choose{2}}}$:
\be
\label{eq:m_mtc}
\tilde{\mathbf M}_{i\ell(j_1,j_2)}= \mathbf M_{i,j_1}\cdot \mathbf M_{i,j_2}, \quad \forall j_1\neq j_2,\textrm{with } j_1,j_2\in \mathcal F,
\ee
where $\ell(j_1,j_2)$ is an index enumerating all distinct unordered pairs $({j_1},{j_2})$. The idea is to construct an `artificial' fault $\tilde f_{\ell(j_1,j_2)}$ and decide that node $i$ is sensitive to it (i.e., $\tilde{\mathbf M}_{i\ell(j_1,j_2)}=1$) iff only one of the faults happening at ${j_1}$ or ${j_2}$ is detected by node $i$. Replacing $\mathbf M$ with $\tilde{\mathbf M}$, defined as in \eqref{eq:m_mtc}, in \eqref{eq:mi_all} leads to the \emph{minimum test cover (MTC)} procedure which maximizes fault isolation performance. 

While other approaches exist in the literature, the MSC and MTC procedures presented above are representative in that they highlight some common issues which lead to a degradation of the subsequent FDI mechanism:
\begin{enumerate}[label=\roman*)]
\item Arguably, the application of a threshold as in \eqref{eq:binarization} discards potentially useful information.
\item The sensor placement procedures are usually either model or data-based. Hybrid ones which make use of both are rarely encountered.
\end{enumerate}
In the following subsection we propose an iterative method which combines unadulterated data (measured/simulated residual values corresponding to multiple fault occurrences) with model information (the graph structure of the network) to decide on an optimal sensor selection. 

\subsection{Graph-aware Gram Schmidt procedure}

Recalling that $\I\subset \{1\dots N\}$ denotes the collection of sensor nodes indices, we note that the submatrix $\bm{R}_\I \in \R^{s\times D}$ will be the only data available for performing FDI.
Thus, we want the low-rank $\bm{R}_\I$ matrix
to approximate as best as possible the full-rank matrix $\bm{R}$. Further, if we look at sensor placement as an iterative process,
then for each new sensor that we place
we get access to the contents of one new row from $\bm{R}$.

Let $\bm{r}^\top_i$ denote row $i$ of matrix $\bm{R}$.
In order to achieve good matrix approximation
we want to make sure that,
when placing a new sensor in node $i$,
the new row $\bm{r}^\top_{i}$ contains as much new information as
possible about the water network.
In other words, we want the projection of $\bm{r}^\top_i$
on the currently selected rows $\bm{R}_\I$ to be minimal:
\be
\begin{aligned}
& i = \underset{j \notin \I}{\arg\min}
& & \norm{\text{proj}_{\bm{R}_\I} \bm{r}^\top_j}_2.
\end{aligned}
\label{proj}
\ee
In this context,
the entire iterative process can be seen as a modified
Gram-Schmidt orthogonalization process
where we create a sequence of $s$ orthogonal vectors
chosen from a set of $N$, selected as in \eref{proj}.

While the process induced by \eqref{proj} might be good enough for matrix approximation,
ignoring the water network's structure (even if implicitly present in the numerical data gathered in the residual matrix $\mathbf R$) is suboptimal.


Considering the underlying undirected weighted graph (via its Laplacian matrix),
we are able to compute the shortest path between all nodes
using Dijkstra's algorithm \cite{Dijkstra59}.
Thus,
we update \eref{proj}
to take into consideration
the distances from the candidate node to the nodes from the existing set
to encourage a better sensor placement spread across the network.

Let $\bm{\delta}_{\I,j} \in \R^{\abs{\I}}$
be the vector whose elements represent the distance between
node $j$ and each of the nodes from set $\I$.
The penalized row selection criteria becomes
\be
\begin{aligned}
& i = \underset{j \notin \I}{\arg\min}
& & \norm{\text{proj}_{\bm{R}_\I} \bm{r}^\top_j}_2 +
\lambda 
\sum_{i \in \I}\frac{1}{\delta_{i,j}}
\label{projdist}
\end{aligned}
\ee
where $\lambda \in \R$ is a scaling parameter.
For $\lambda = 0$ \eref{projdist} is equivalent to \eref{proj}.

The penalty mechanism works as follows:
if the projection is small
and the sum of distances from node $j$ to the nodes of $\I$ is large,
then the distance penalty is small also and node $j$ is a good candidate.
On the other hand,
if the sum of distances is small
then the penalty grows and the possibility of selecting $j$ decreases.

The result is a data topology-aware selection process,
that encourages a good distribution of sensors inside the network
in order to facilitate FDI. 
We gather the instructions necessary for sensor placement in
Algorithm \ref{alg:placement}.

\begin{algorithm2e}[!ht]
\KwData{\hspace{1mm}training residuals $\bm{R} \in \R^{N \times D}$, \\
	\hspace{13mm}shortest path between nodes
	  $\bm{\Delta} \in \R^{N \times N}$, \\
	\hspace{13mm}sensors $s \in \N$, \\
	\hspace{13mm}distance penalty $\lambda \in \R$}
\KwResult{sensor nodes $\I$}
\BlankLine

Find dominant row:
  $i = \underset{k}{\arg\max}\norm{\bm{r}_k^\top}_2$, $1 \le k \le N$ \\
Initial set: $\I = \{i\}$ \\
Orthogonal rows: $\bm{U} =
	\begin{bmatrix}
		\frac{\bm{r}^\top_i}{\norm{\bm{r}^\top_i}}
	\end{bmatrix}$ \\

\For{$k = 2$ \KwTo $s$}{
  Compute inner-products $\bm{P} = \bm{R}_{\I^c}(\bm{R}_{\I})^\top$ \\
  Project on the selection sub-space: $\bm{S} = \bm{P}\bm{U}^\top$\\
  Compute projection norms: $\bm{n} =
    \begin{bmatrix}
      \norm{\bm{s}_1^\top}_2 & \norm{\bm{s}_2^\top}_2 & \dots &
      \norm{\bm{s}_{n-k}^\top}_2
    \end{bmatrix}$\\
  Distance penalty:
  $n_j = n_j + \lambda\sum_{i \in \I}\delta_{i,j}^{-1}$,
    $j \in \I^c$ \\
  $i = \arg\min n_j$, $j \in \I^c$ \\
  $\I = \I \cup \{i\}$ \\
  $\bm{u} = \bm{r}^\top_i - \bm{s}^\top_i$ \\
  $\bm{U} =
	\begin{bmatrix}
		\bm{U} & \frac{\bm{u}}{\norm{\bm{u}}}
	\end{bmatrix}$ \\
}
\caption{Gram-Schmidt Sensor Placement}
\label{alg:placement}
\end{algorithm2e}
First,
we select the row whose energy is largest (step 1)
and place the first sensor there (step 2).
We place the normalized row in the first column of matrix $\bm{U}$ (step 3)
where we will continue to store the orthogonal vector sequence
as discussed around \eref{proj}.
This auxiliary matrix helps us with future projections computations.
From this initial state,
step 4 loops until we place the remaining sensors.
We begin iteration $k$
by projecting the candidate rows $\I^c$ onto the existing selection (step 5).
The $p_{i,j}$ element represents the projection of the candidate row $i$ on the
selected node $j$.
Step 6 completes the projection by multiplying the inner-products with the
corresponding orthogonal vectors.
These two steps are required by
\eref{proj} to compute the projection of all $r_j^T$ on the selected rows ${\cal I}$.
Next, we store in vector $\bm{n}$ the projection norm of each candidate
row (step 7).
We are given the shortest path between any two nodes in matrix $\bm{\Delta}$,
where $\delta_{i,j}$ represents the distance from node $i$ to node $j$.
In step 8, we penalize the projections by summing up the inverse of the distance
from candidate node $j$ to each of the selected nodes.
Node $i$ corresponding to the smallest element in $\bm{n}$ is found (step 9)
and added to the set $\I$.
Steps 11 and 12 perform the Gram-Schmidt orthogonalization process:
first the redundant information is removed from the row $i$
(by substracting the projection on the old set)
and then the resulting vector is normalized and added to the orthogonal matrix
$\bm{U}$.

\begin{rem}
The algorithm computations are dominated by
the large matrix multiplications in steps 5 and 6.
At step $k$, we need to perform $2Dk(N-k)$ operations
in order to obtain the matrix $\bm{S}$.
The rest of the instructions require minor computational efforts in comparison.
This results in a complexity of $O(s^2ND)$ for the entire loop.\eor
\end{rem}
\begin{rem}
Arguably, the weights appearing in the Laplacian graph should be proportional with the headloss between two linked nodes. This is not trivial since the headloss depends nonlinearly on pipe length, diameter and roughness coefficient, see \eqref{eq:hw}. \eor
\end{rem}
\subsection*{Sensor placement in the Hanoi network}
\begin{figure}[ht!]
\centering
\includegraphics[width=\columnwidth]{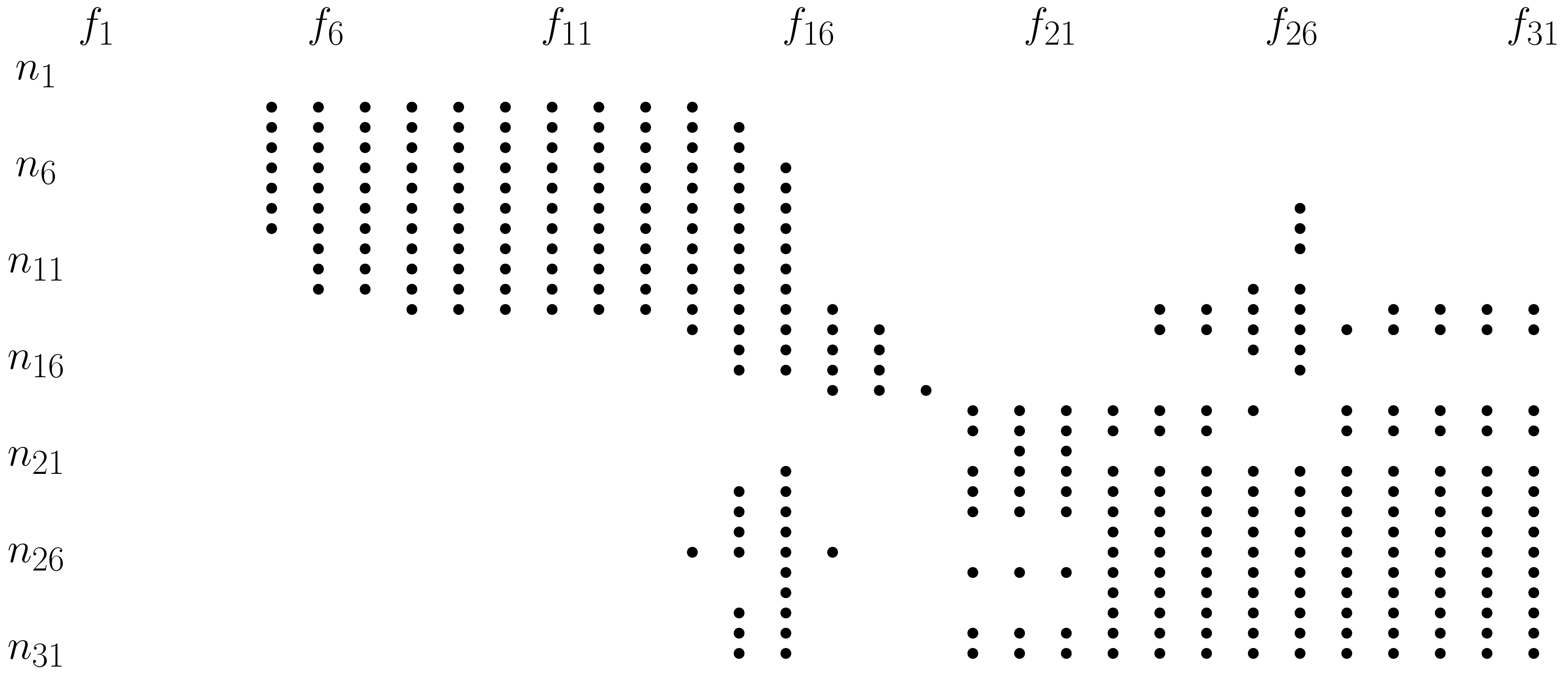}
\caption{Illustration of the fault signature matrix for the Hanoi water network.}
\label{fig:fault_signature_matrix}
\end{figure}

Using the example from \secref{sec:prelim}, we now consider the three methods introduced earlier (MSC, MTC and Graph-GS) to generate sensor placements. We limit to the nominal profile case and compute the fault signature matrix $\mathbf M$ as in \eqref{eq:binarization}, for $\tau=3$. The result is illustrated in \figref{fig:fault_signature_matrix} where a bullet at coordinates $(i,j)$ means that the i-th node detects the j-th fault for at least one of its magnitudes. Arguably, instead of considering ``any'' we may consider that ``all'' or ``most'' of the fault magnitudes have to be detected at the node in order to consider it ``fault-affected''. Ultimately, classifying a node as being fault-affected is a mater of choice depending on the problem's particularities.

Applying the MSC procedure as in \eqref{eq:mi_all} leads to the sensor selection $\{1,13,20,26,31\}$. Constructing the extended signature matrix $\tilde{\mathbf M}$ as in \eqref{eq:m_mtc} and using it for the MTC procedure leads to the sensor selection $\{6,16,17,19,24\}$. Lastly, the Graph-GS approach retrieves the selection $\{1,2,3,10,28\}$ for the parameter $\lambda=10^4$. In all cases, we assumed $s=5$ (note that the MSC/MTC procedures may select fewer than $s$ nodes since \eqref{eq:mi_c} is an inequality). 

The MSC and MTC procedures are able to run for this proof-of-concept network but the required number of binary variables increases in lock-step with the number of junction nodes for MSC and exponentially for MTC (e.g., in this particular example, 31 and, respectively, ${31 \choose 2}=465$). The computation times are negligible here but they increase significantly in the case of large systems (as further seen in \secref{sec:sims}). Lastly, by counting the cases for which $\gamma_j=0$ from \eqref{eq:mi_all} we estimate the number of fault detection errors in MSC (3 cases) and of fault isolation errors in MTC (34 cases). On the other hand, the Graph-GS procedure is much less sensitive to problem size and can handle easily large problems.

\begin{figure}[ht!]
\centering
\includegraphics[width=\columnwidth]{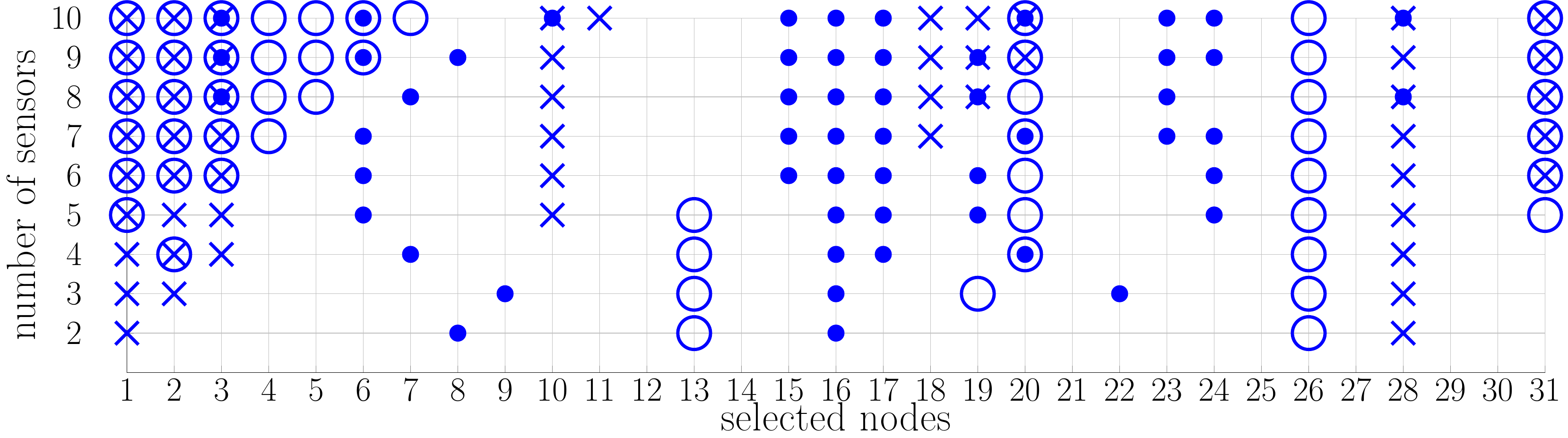}
\caption{Illustration of sensor selection for various methods.}
\label{fig:sensor_placement}
\end{figure}

\figref{fig:sensor_placement} illustrates the selections resulted for each method (circle, bullet and 'X' symbols for MTC, MSC and Graph-GS) for sensor numbers ranging from 2 to 10.

A word of caution is in order: regardless of method, the performance of a sensor selection is not truly validated until the FDI block is run and its output is compared with the actual fault occurrences. This is the scope of \secref{sec:dictionary}. 

\section{Dictionary Learning and Classification Strategies}
\label{sec:dictionary}

Dictionary learning (DL) \cite{DL_book} is an active field
in the signal processing community
with multiple applications such as
denoising,
compression,
super-resolution,
and classification.
Recent studies have also shown good results when dealing with anomaly detection in general~\cite{BPI20_ifac, IPA19_amlsoa, IB10_itwist},
and particularly when applied to the FDI problem in water networks~\cite{IS17_ifac,SI19_mpdl}.

\subsubsection*{Dictionary Learning}
Starting from a sample
$\bm{Y} \in \R^{s\times D}$, our aim is to find an overcomplete base
$\bm{D} \in \R^{s\times B}$,
called the dictionary,
with whom we can represent the data by using only a few of its columns,
also called atoms.
Thus, we express the DL problem as
\bse
\begin{align}
 \underset{\bm{D}, \bm{X}}{\min} &  \norm{\bm{Y}-\bm{DX}}_F^2 \\
\label{eq:dl_spasity}\text{s.t.} &\norm{\bm{x}_\ell}_{0} \leq s_0,\ \ell = 1:D \\
\label{eq:dl_norm}&\norm{\bm{d}_j} = 1, \ j = 1:B,
\end{align}
\label{dict_learn}
\ese
where $\bm{X} \in \R^{B \times D}$ are the sparse representations
corresponding of the $\bm{Y}$ signals.
\eqref{eq:dl_spasity} dictates that
each column $\bm{y}_\ell \in \bm{Y}$
has a sparse represention $\bm{x}_\ell$
that uses at most $s$ atoms from $\bm{D}$
(i.e. $\bm{y}_\ell$ is modeled as
the linear combination of at most $s_0$ columns from $\bm{D}$).
\eqref{eq:dl_norm} is there to avoid the multiplication ambiguity
between $\bm{D}$ and $\bm{X}$ (i.e., it allows to interpret the atoms as directions
that the sparse representations follow. Thus, the elements in $\bm{X}$ act as scaling coefficients of these directions).

\begin{rem}
Solving \eref{dict_learn} is difficult because the objective is non-convex and
NP-hard.
Existing methods approach the problem through
iterative alternate optimization techniques.
First the dictionary $\bm{D}$ is fixed and we find the representations $\bm{X}$,
this is called the sparse representation phase
and is usually solved by greedy algorithms
among which Orthogonal Matching Pursuit (OMP) \cite{PRK93omp}
is a fast, performant and popular solution.
Next,
we fix the representations and we find the dictionary.
This is called the dictionary training or dictionary refinement phase
and most algorithms solve it by
updating each atom at a time
(and sometimes also the representations using it)
while fixing the rest of the dictionary.
Popular routines are
K-SVD \cite{AEB06}
and Approximate K-SVD (AK-SVD)\cite{RZE08}.
A few iterations of the representation and dictionary training steps
usually bring us to a good solution. \eor
\end{rem}

\subsubsection*{Dictionary classification}

The choice of the $s$ nonzero elements of a given column $\bm{x}_\ell$ (also called the representation support), highlights the participating atoms. These form a specific pattern which allows us to emit certain statements about the data which led to it.
For example,
let us assume that we can split the input data $\mathbf{Y}$ into distinct classes.
Then, it follows naturally that signals from a certain class will probably
use a characteristic pattern more than the signals from the other classes. 
Thus, we can classify a signal as being part of certain classes by just looking at the atoms used in its representation.
In its direct form \eqref{dict_learn}, the dictionary learning and classification suffers (at least from the viewpoint of FDI analysis) from a couple of shortcomings:
\begin{enumerate}[label=\roman*)]
\item\label{item:issue_1} the procedure is not discriminative: the learned dictionary may in fact lead to classifications with overlapping patterns of atom selection;
\item\label{item:issue_2} problem dimensions can grow fast as the number of network nodes increases.
\end{enumerate}

Let $c$ be the number of classes\footnote{Given that the data-items in $\bm{Y}$ are already labeled, we already know to which class they belong to.}
and let us assume, without any loss of generality,
that $\bm{Y}$ is sorted and can be split into $c$ sub-matrices,
each containing the data-items of a single class:
$\bm{Y} = [\bm{Y}_1 \ \bm{Y}_2 \ \ldots \ \bm{Y}_c]$.
An alternative method to \eqref{dict_learn}
called
Label consistent K-SVD (LC-KSVD) \cite{JLD13}
adds extra penalties to \eref{dict_learn}
such that the
atoms inherit class discriminative properties
and, at the same time,
trains a classifier $\bm{W}$ to be used afterwards,
together with $\bm{D}$,
to perform classification.

Let $\bm{H} \in \R^{c\times D}$ be the data labeling matrix 
with its columns corresponding to the ones in $\bm{Y}$.
If $\bm{y}_j$ belongs to class $i$
then $\bm{h}_j = \bm{e}_i$,
where $\bm{e}_i$ is the i-th column of the identity matrix.
Let matrix $\bm{Q} \in \R^{B\times D}$ be the discriminative matrix
whose row numbers correspond to the dictionary atoms and
whose column numbers correspond to the training signals.
Column $\bm{q}_j$ has ones in the positions corresponding to the atoms
associated with the class that $\bm{y}_j$ belongs to and zeros elsewhere.
Usually atoms are split equally among classes.
When the training signals are sorted in class order,
matrix $\bm{Q}$ consists of rectangular blocks of ones arranged diagonally.
LC-KSVD solves the optimization problem
\be
\underset{\bm{D},\bm{X},\bm{W},\bm{A}}{\min}
  \norm{\bm{Y}-\bm{DX}}_F^2
    + \alpha \norm{\bm{H}-\bm{WX}}_F^2
    + \beta \norm{\bm{Q}-\bm{AX}}_F^2
\label{lcksvd_dl}
\ee
where the first term is the DL objective \eref{dict_learn}.
The second term connects the label matrix $\bm{H}$
to the sparse representations $\bm{X}$
through matrix $\bm{W}$.
We can view this as a separate DL problem
where the training data are the labels and
the dictionary is in fact a classifier.
The trade-off between small representation error
and accurate classification is tuned by parameter $\alpha$.
The third term learns $\bm{A}$ such that
the linear transformation of $\bm{X}$ approximates $\bm{Q}$.
Depending on the way we built $\bm{Q}$ (see above),
this enforces that, first, only a few number of atoms can characterize a class and, second,
these atom combinations can not appear and are different from those for other classes.

After the learning process is over,
in order to classify a signal $\bm{y}$,
we need to first compute its representation with dictionary $\bm{D}$,
again by using OMP or a similar algorithm,
and then find the largest entry of $\bm{Wx}$
whose position $j$ corresponds to the
class that $\bm{y}$ belongs to
\be
j = \underset{i=1:c}{\arg\max} (\bm{Wx})_i.
\label{class_discrim}
\ee

\begin{rem}
While \eqref{lcksvd_dl} introduces additional variables it is, qualitatively, similar with \eqref{dict_learn} as it can be reduced to a similar formulation. Indeed, \eref{lcksvd_dl} can be reformulated as a ``composite dictionary learning problem''
\be
\underset{\bm{D},\bm{X},\bm{W},\bm{A}}{\min}
\norm{\baa{c} \bm{Y} \\ \sqrt{\alpha} \bm{H} \\ \sqrt{\beta} \bm{Q} \eaa -
\baa{c} \bm{D} \\ \sqrt{\alpha} \bm{W} \\ \sqrt{\beta} \bm{A} \eaa \bm{X}}_F^2,
\label{lcksvd_ksvd}
\ee
where $\mathbf D, \mathbf W, \mathbf A$ are learned from data provided by $\mathbf Y, \mathbf H$ and $\mathbf Q$, respectively.
Note that in this particular instance, after the learning process $\bm{A}$ is discarded
as it indirectly instilled discriminative properties to dictionary $\bm{D}$.\eor
\end{rem}


\subsection{Online learning}
\label{sec:dl-online}

When dealing with large scale distribution networks
the problem dimensions explode.
Typical water networks may have hundreds or even thousands of nodes. Take as an example the case of a network with $5,000$ nodes.
If each node represents one class,
with 3 atoms per class we end up with a 15,000 column dictionary.
Training on 30,000 signals,
we end up with a $15,000\times30,000$ representations matrix.
The computations involved become prohibitive on most systems.
To accommodate large-scale scenarios
we propose an online learning alternative.

Online DL handles one signal at a time,
thus most operations become simple vector multiplications.
At time $t$, we are given signal $\bm{y}$
which we use to update the current dictionaries $\bm{D}^{(t)}$,
$\bm{W}^{(t)}$ and $\bm{Q}^{(t)}$.
In order to speed-up results,
a pre-training phase can be ensured
where \eref{lcksvd_dl} is performed
on a small batch of signals.
The TODDLeR algorithm~\cite{IB19_toddler}
adapts objective \eref{lcksvd_dl} for online learning
using the recursive least squares approach~\cite{SE10_rls}.
Besides signal classification,
its goal is to learn from all incoming signals: labeled or not.
TODDLeR ensures that the model is not broken through miss-classification by regulating the rate of change each signal brings
\be
\begin{aligned}
\min_{\bm{D}, \bm{x}, \bm{W}, \bm{A}} &\|\bm{y}-\bm{Dx}\|_{2}^2
+ \alpha \|\bm{h}-\bm{Wx}\|_{2}^2 + \beta \|\bm{q}-\bm{Ax}\|_{2}^2 \\
& + \lambda_1\norm{\bm{W}-\bm{W}_0}^2_F + \lambda_2\norm{\bm{A}-\bm{A}_0}^2_F.
\end{aligned}
\label{opt_toddler}
\ee
For convenience, we dropped the $(t)$ superscripts above.
The problem does not have a closed-form solution and is solved in two steps.
First, we solve the \eref{lcksvd_dl} problem for a single vector
using the first three terms in \eref{opt_toddler}.
As shown in \cite{IB19_toddler},
this translates to updating $\bm{D}$, $\bm{W}$, $\bm{A}$
through a simple rank-1 update 
based on the $\bm{y}$ and its sparse representation $\bm{x}$.
Keeping everything fixed in \eref{opt_toddler}
except for $\bm{W}$ and $\bm{A}$ respectively,
leads to the following two objectives
\be
f(\bm{W}) = \norm{\bm{h}-\bm{Wx}}_{2}^2 + \lambda_1\norm{\bm{W}-\bm{W}_0}^2_F
\label{opt_W}
\ee
\be
g(\bm{A}) = \norm{\bm{q}-\bm{Ax}}_{2}^2 + \lambda_2\norm{\bm{A}-\bm{A}_0}^2_F
\label{opt_A}
\ee
meant to temper the updates brought by $\bm{y}$ to the dictionaries in the first step.
Equations \eref{opt_W} and \eref{opt_A} are
simple least-squares problems.
By looking at $f$ and $g$ as generalized Tikhonov regularizations,
it was motivated in~\cite{IB19_toddler} that good parameter choices are
$\lambda_{1,2}=\norm{\bm{G}}_2$
or
$\lambda_1=\norm{\bm{W}_0}_2$, $\lambda_2=\norm{\bm{A}_0}_2$.
Here $\bm{G}=\bm{XX}^T$ is the Gram
matrix of the pre-train representations
that is then rank-1 updated in the online phase
by each incoming signal $\bm{G} + \bm{xx}^T$.

\subsection{FDI mechanism}

Recall that the DL procedure classifies an input vector wrt a set of a priori defined classes. Thus, assimilating a fault to a class means that the classification step of the DL procedure actually implements fault detection and isolation.  The details are provided in \algref{alg:fdi}.



\begin{algorithm2e}[!ht]
\KwData{\hspace{1mm}restriction of the residual matrix $\bm{R}_\I \in \R^{s \times D}$}
\KwResult{estimated class set $\hat{\mathcal L}$}
\BlankLine

partition indices into pre-train, train and test: $\I_{pt}\cup \I_{tr}\cup \I_{te}=\{1\dots D\}$, $\I_{pt}\cap \I_{tr}=\emptyset$, $\I_{pt}\cap \I_{te}=\emptyset$, $\I_{tr}\cap \I_{te}=\emptyset$\\
pre-train $D,A,W$ as in \eqref{lcksvd_dl} for labelled data $\mathbf R_\I^{\I_{pt}}$\\
\For{$r\in \mathbf R_\I^{\I_{pt}}$}{
  update online $D,A,W$ as in \eqref{opt_toddler} for labelled residual vector $r\in \mathbf R_\I^{\I_{pt}}$\\
}
init the estimated class collection: $\hat{\mathcal L}=\emptyset$\\
\For{$r\in \mathbf R_\I^{\I_{te}}$}{
  update online $D,A,W$ as in \eqref{opt_toddler} for unlabelled residual vector $r\in \mathbf R_\I^{\I_{te}}$\\
  retrieve estimated active class $\hat\ell$ as in \eqref{class_discrim}\\
  update the estimated class set $\hat{\mathcal L}=\hat{\mathcal L}\cup \{\hat\ell\}$\\
}
\caption{FDI Mechanism}
\label{alg:fdi}
\end{algorithm2e}

Considering the residual matrix $\mathbf R$ from \eqref{eq:resmat} and the sensor selection $\I$ obtained in \secref{sec:placement}, we arrive to sub-matrix $\mathbf R_{\I}$. To this we associate the fault labels $\mathcal L$ (the active fault index for each of the columns of $\mathbf R_{\I}$).

Step 1 of the algorithm divides the residuals and associated labels into disjoint `pre-train', `train' and `test' collections $\I_{pt}$, $\I_{tr}$, $\I_{te}$. These are used in steps 2, 4 and 8 to construct and respectively update the dictionary. Step 9 handles the actual FDI procedure by selecting the class best fitted to the current test residual ($r\in \mathbf R_\I^{\I_{te}}$). The class estimations are collected in $\hat{\mathcal L}$, at step 10 and compared with the actual test fault labels\footnote{The test fault labels are available since the analysis is carried out in simulation, in reality, the residuals obtained at runtime do not have such a label.} ${\mathcal L}^{\I_{te}}$ to assess the success criterion of the FDI mechanism
$S=(|\hat{\mathcal L}\cap \mathcal L^{\I_{te}}|/|\mathcal L^{\I_{te}}|)\cdot 100$.

\begin{rem}
\label{rem:neighbors}
Arguably, it makes sense tweaking criterion $S$  to count for near misses: the classification is successful not only if the correct node is identified but also if one of its neighbors is returned by the classification procedure. \eor
\end{rem}
\begin{rem}
\label{rem:confidence}
By construction, \eqref{class_discrim} returns the index corresponding to the largest value in the vector $\mathbf W\mathbf x$. This ignores the relative ranking of the classifiers, as it does not attach a degree of confidence for the selected index (i.e., large if there is a clear demarcation between classifications and small if the values are closely grouped. \eor
\end{rem}
\subsection*{Illustration of the FDI mechanism}

In our experiments each network node represents one class. During DL we used an extra shared dictionary to eliminate the commonalities within class-specific atoms~\cite{DL_book}.
This lead to $c=32$ classes for which we allocated 3 atoms per class
leading to $n=3c$ dictionary atoms.
An initial dictionary was obtained through pre-training on 2480 signals.
Afterwards we performed online training on
2170 signals.
Cross-validation showed good results when using $\alpha=4$ and $\beta=16$.
When updating $\bm{W}$ and $\bm{A}$ we used
$\lambda_{1,2}=\norm{\bm{G}}_2$.
With the resulting dictionaries we tested the FDI mechanism online
on 4650 unlabeled signals. Applying the Graph-GS method for sensor selection, the rate of success was $S=80.09\%$.

\begin{figure}[ht!]
\centering
\includegraphics[width=.85\columnwidth]{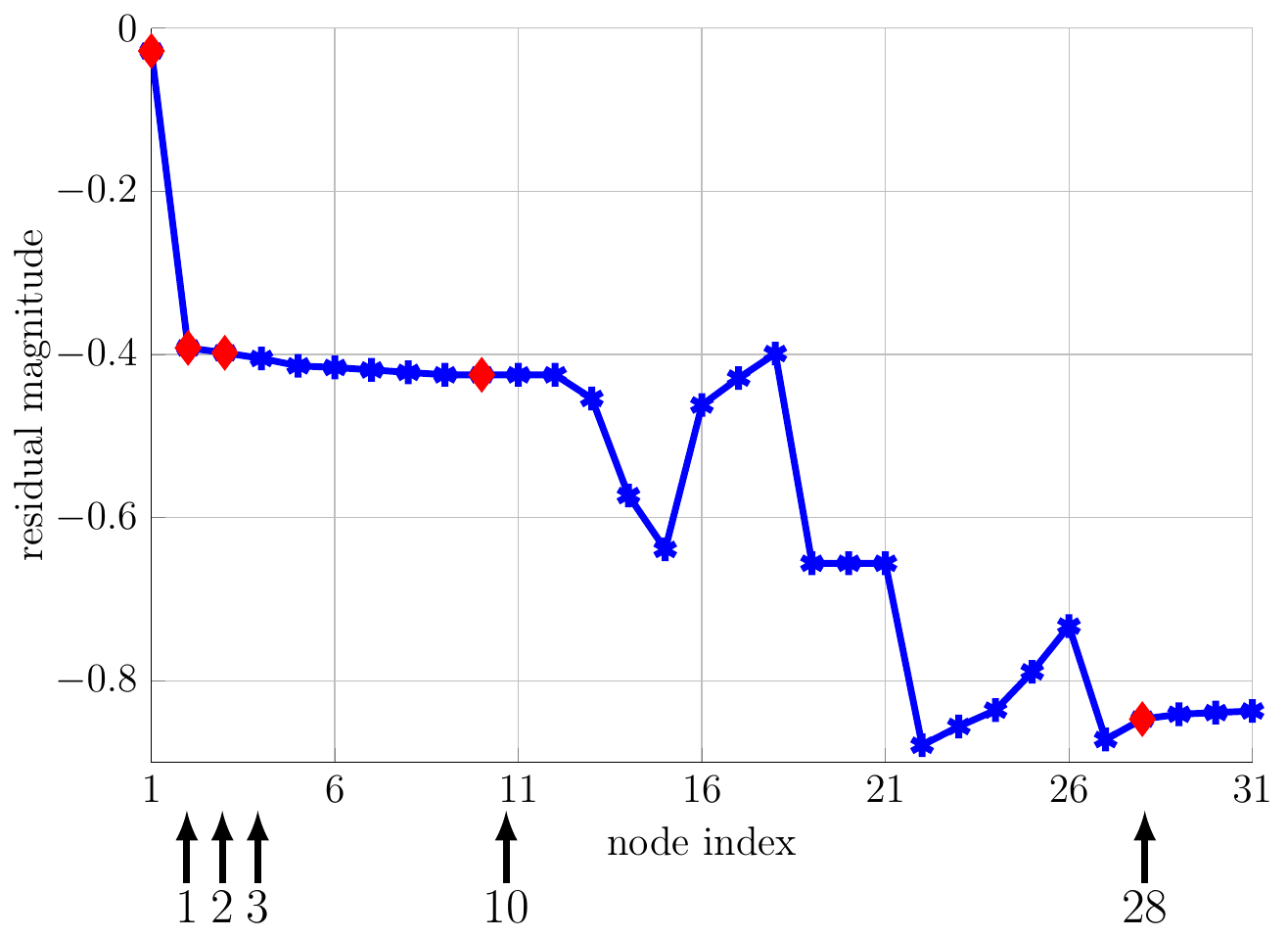}
\caption{Illustration of fault detection and isolation.}
\label{fig:fdi_dl}
\end{figure}

For illustration, we show in \figref{fig:fdi_dl} the full (blue line with bullet markers) residual signal corresponding to class $22$ (i.e., the case where 22th node is under fault) and the actually-used data (red diamond markers), at nodes with sensors (those with indices from $\{1,2,3,10,28\}$).
The actual classification was done as in \eqref{class_discrim} and resulted in a classifier vector $\mathbf x$ whose non-zero values are $\{7e-6, -1.2e-4,-7e-5,0.99, 1e-3\}$, at indices $\{45, 46, 52, 64, 93\}$. Clearly, the most relevant atom in the description is the one with index $64$ which lies in the subset $\{63,64,65\}$ corresponding to class $22$. The classification $\bm{Wx}$ produces an indicator vector where the first and second largest values are $0.999$ and $1e-5$ thus showing that the procedure unambiguously produces the correct response (see \remref{rem:confidence}).

Further, we consider not only the success rate as defined in step 12 of \algref{alg:fdi} but also count the cases where the fault is identified in the correct node's neighbors and in the neighbors' neighbors (as per \remref{rem:neighbors}). This leads to an increase from $80,09\%$ to $90.69\%$ and $98.92\%$, respectively, for the success rate.

Lastly, using the MSC sensor selection procedure we arrive to success rates $60.95\%$, $78.11\%$ and $88.56\%$ which proved to be significantly lower than the Graph-GS selection method. The MTC method, even for this small-scale network does not provide a solution.

Direct comparisons of results are somewhat difficult even when using the same network due to differences in flow profile, leakage values and the presence or absence of disturbances. As a qualitative indication of our relative performance we compare with the results of \cite{casillas2013optimal}, that uses a well-accepted approach in model-based leak localization. Using a performance indicator (defined by eqs. (14, 17,18) in \cite{casillas2013optimal}) which increases linearly with the distance (in the topological sense) between the actual node under fault and the estimated one, we arrive at a value of $0.453$, larger than the results shown, e.g., in Table 3 of \cite{casillas2013optimal} which range from $0.02$ to $0.157$. The results are reasonably close if we take into account our more realistic setup:  more fault magnitudes (and some of them significantly smaller and hence harder to detect than the ones used in \cite{casillas2013optimal}) and the presence of noise-affected flow profiles (in \cite{casillas2013optimal} only a nominal profile is used).  Note that the approach in \cite{casillas2013optimal} is using the hydraulic model of the network for leak localization, thus requiring much more information than the data-driven method proposed here. So, these similar leak localization results makes the proposed method competitive against model-based methods without requiring a detailed and well-calibrated model of the network.

From a computational viewpoint, our method is significantly better in the sensor selection step (\cite{casillas2013optimal} executes a semi-exhaustive search through a genetic algorithm whereas we solve an almost-instantaneous Gram-Schimidt procedure) and comparable in FDI performance (for us a quadratic optimization problem, for \cite{casillas2013optimal} a sensitivity matrix computation).

Further, 
we compare our FDI method with the standard linear learning methods from \cite{soldevila2017leak}. We do this both to compare with state-of-the art classifiers and because the methods from \cite{casillas2013optimal} do not scale reasonably for large-scale networks.


\section{Validation over a generic large-scale water network \protect\footnotemark }
\label{sec:sims}

\footnotetext{Code available at \url{https://github.com/pirofti/ddnet-online}}

\begin{figure}[!ht]
    \centering
    \includegraphics[width=\columnwidth]{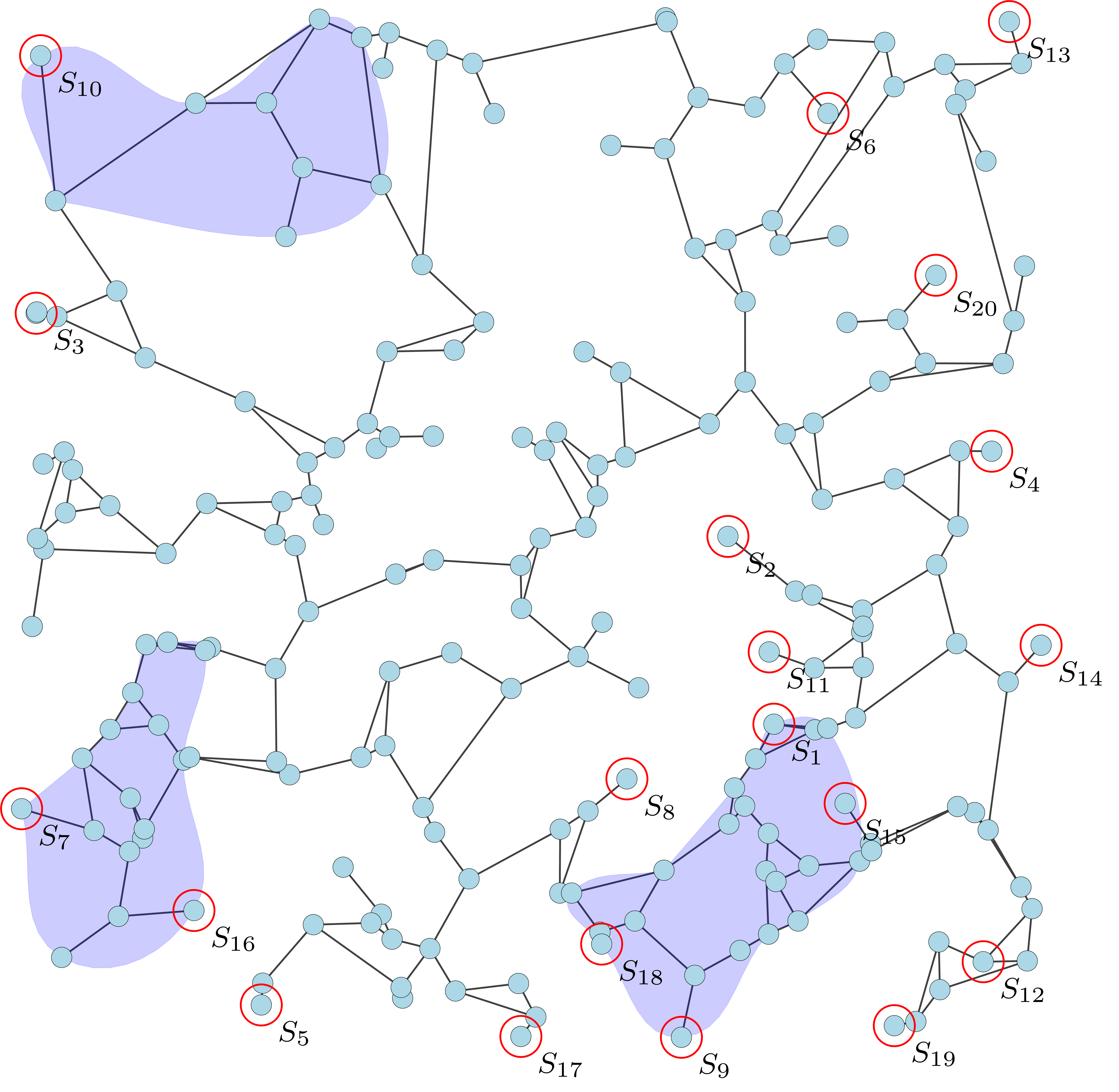}
    \caption{Large-scale generic water network}
    \label{fig:network_generic}
\end{figure}

To illustrate the DL-FDI mechanism, we test it over a large-scale generic network obtained via the EPANET plugin WaterNetGen \cite{muranho2012waternetgen}. To generate the network we considered 200 junctions, 1 tank and 253 pipes interconnecting them into a single cluster, as shown in \figref{fig:network_generic}.

\subsection{Setup}
To test our algorithms, we first take a nominal profile of node demands and perturb it with $\pm 5\%$ around its nominal values. Further, we consider that fault events are denoted by non-zero emitter\footnote{In the EPANET software, to each junction node corresponds an emitter value which denotes the magnitude of the node leakage.} values in the EPANET emulator. With the notation from \algref{alg:fdi} we run the EPANET software to obtain residual vectors (in absolute form) as follows:
\begin{enumerate}[label=\roman*),nosep]
    \item 2400 = 200 $\times$ 12 pre-train residuals; under the nominal profile, for each fault event we consider emitter values from the set of even values $\{8,10,\dots,30\}$;
    \item 2400 = 200 $\times$ 12 train residuals; for each node we consider 12 random combination of profile (from the set $\{1,2,\dots, 10\}$) and of emitter value (from the set of odd values $\{9,11,\dots,31\}$);
    \item 3200 test residuals; we select random combinations of profile, fault and emitter values (taken from the sets $\{1,2,\dots, 10\}$, $\{1,2,\dots, 200\}$ and $\{1,2,\dots, 31\}$, respectively).
\end{enumerate}
For further use we also divide the graph into $17$ communities using the community detection tool \cite{blondel2008fast}. For illustration we depict in \figref{fig:network_generic} three of these (semi-transparent blue blobs).

The first step is to apply the sensor placement \algref{alg:placement} to retrieve the sub-matrix $R_{pt}$ which gathers the pre-train residuals, taken at indices corresponding to sensor placements. The result is visible in \figref{fig:network_generic} where we plotted (red circles) the first $20$ sensor selections. Note that, due to the particularities of the Graph-GS method, each lower-order sensor selection is completely included within any larger-order sensor selection, e.g., selecting $5$ sensors gives the collection $\{14,21,135,142,192\}$ which is a subset of  $\{14,21,41,113,117,135,137,142,170,192\}$, obtained when selecting $10$ sensors.

As a first validation we consider that each node fault is a distinct class and apply the DL-FDI mechanism described in \algref{alg:fdi} to detect and isolate them. We quantify the success of the scheme in three ways, by counting all the cases where: 
\begin{enumerate}[label=S\arabic*),nosep]
    \item the estimated node equals the actual node under fault;
    \item the estimated node is, at most, the neighbor of the node under fault\footnote{Arguably this criterion imposes no performance penalty. The fault event is in fact a pipe leakage and associating the fault with a node is a simplification usually taken in the state of the art. In reality, if a node is labelled as being faulty, the surrounding ones need to be checked anyway.};
    \item the estimated node is, at most, the once-removed neighbour of the node under fault.
\end{enumerate}   The three previous criteria can be interpreted as 0, 1 and 2-distances in the network's Laplacian. Arbitrary, n-distance, neighbors can be considered but their relevance becomes progressively less important.

The aforementioned community partitioning is another way of solving the FDI problem: each class corresponds to a community, i.e., any fault within the the community is labelled as being part of the same class. This approach leads to an additional success criterion:
\begin{enumerate}[label=S\arabic*),resume,nosep]
\item the estimated class corresponds to the the community within which the fault appears.
\end{enumerate}

In our simulations we set the parameters in \eref{opt_toddler}
as follows:
$\alpha=4$ and 
$\beta=16$
for the classification dictionaries as
found via cross-validation~\cite{IB19_toddler, JLD13}; 
for the update regularization
we initialized $\lambda_{1,2}=8$
and proceeded with
automated parameter tuning $\lambda_{1,2}=\norm{\bm{G}}_2$.
TODDLeR was pre-trained, as earlier described, by first
running LC-KSVD~\cite{JLD13} on a small dataset
in order to obtain an initial dictionary $D$ and representations $X$.
LC-KSVD used 20 iterations of AK-SVD~\cite{RZE08} to train the atoms block belonging to each class and then 50 more iterations on the entire dictionary.

To asses our method we compare it with linear classifiers
such as k-NN, SVM, and Na{\"i}ve-Bayes,
as commonly used for this type of problem (see, e.g., \cite{soldevila2017leak}).
We mention that,
even though these are also linear learning methods,
they operate in bulk,
by processing the entire dataset at once.
This becomes prohibitive for large distribution networks
and it is indeed the reason why we consider here an online method which provides reduced memory footprint and faster execution times.

\subsection{Results}
\begin{figure}[!ht]
    \centering
    \subfloat[DL-FDI method, criterion comparison]{\label{fig:dl_generic_2_success}\includegraphics[width=\columnwidth]{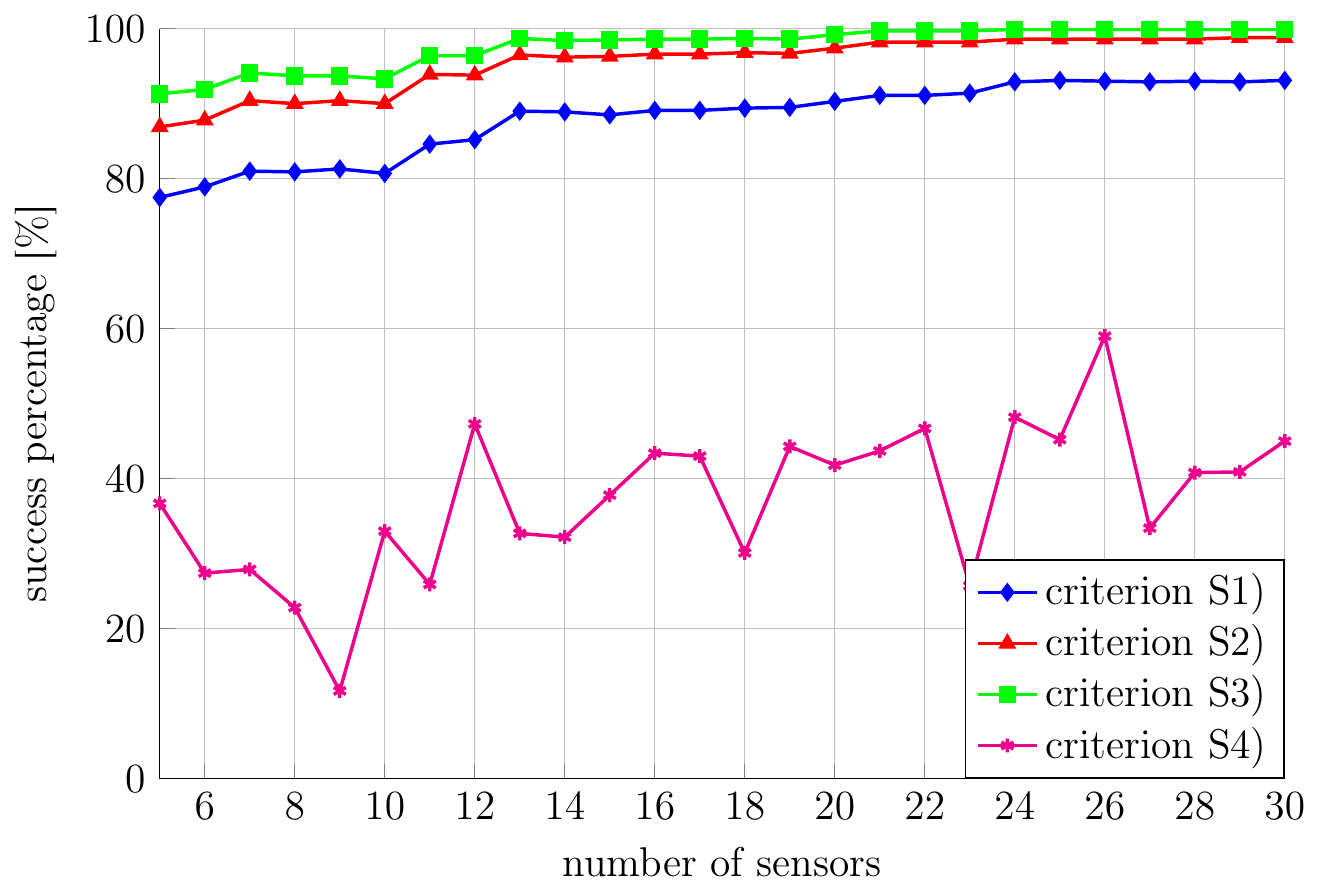}}\\
    \subfloat[DL-FDI versus k-NN and SVM classifiers]{\label{fig:dl_generic_2_success_class}\includegraphics[width=\columnwidth]{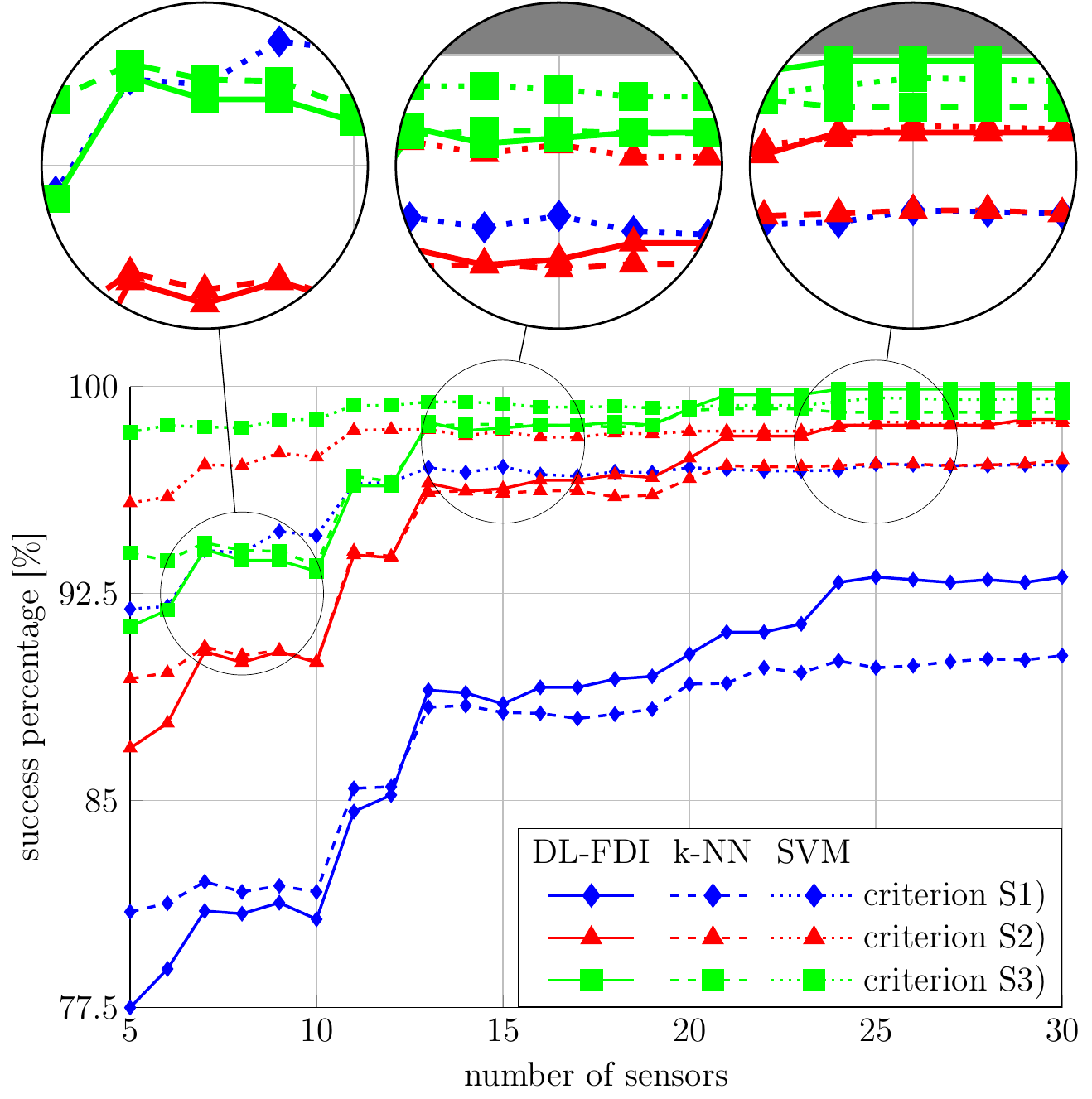}}
    \caption{FDI success rates for a large-scale water network}
    \label{fig:dl_generic_2_success_all}
\end{figure}

Running \algref{alg:fdi} for a number of selected (as in \algref{alg:placement}) sensors ranging from 5 to 30 we obtain the success rates shown in \figref{fig:dl_generic_2_success}. 


Several remarks can be drawn. First, and as expected, an increase in sensors, generally leads to an increase in performance. Still, care should be taken with the numbers considered: we note that even a small number (5) gives a good success rate and that after around $\geq 15$ sensors the performance improvements taper off. Second, the classification failures appear to be `near-misses' as can be seen when comparing the S1), S2) and S3) criteria. The S2) and S3) values approach fast 100\% which means that the difference (in the topological sense) between the estimated and the actual node under fault is small). In fact, having 24 or more sensors selected means that (as illustrated by the S3) criterion) the estimated fault location is never further away than 2 nodes from the actual fault location. Reducing the number of classes as in criterion S4) significantly reduces the computation time but also leads to a marked decrease in performance (which does not appear to improve with an increase in the number of sensors). We consider this to be due to the uniform distribution of the nodes in the synthetic network considered here. We expect that applying this method in a typical residential water network (where dense neighborhoods are linked within the larger network by few long pipes) will provide better results.

To asses our method we compare it with linear classifiers commonly used for this type of problem (see, e.g., \cite{soldevila2017leak}). Specifically, 
we consider the k-NN and SVM classifiers and illustrate the results in \figref{fig:dl_generic_2_success_class}: we consider criterions S1, 2 and 3) with blue diamond, red triangle and green square markers and plot the success rates for our implementation (DL-FDI), k-NN and SVM with solid, dashed and respectively, dotted lines.

Performance-wise, k-NN proves slightly better at small number of sensors but DL-FDI proves superior for larger values. SVM, on the other hand is better for small numbers of sensors and comparable with DL-FDI at larger values. Time-wise, k-NN is significantly more rapid than DL-FDI and SVM is orders of magnitude slower than either of them. 

We note however that both k-NN and SVM consider all training data simultaneously
and that the model they provide is fixed,
whereas DL-FDI is implemented in its online form where residuals from the training and testing steps are fed to the algorithm one by one and the model is continuously updated. This means that k-NN and SVM are more sensitive to problem dimension and that at some point the memory requirements become prohibitive (which is not the case for our online DL-FDI implementation).

We also implemented the Bayesian method proposed in \cite{soldevila2017leak} but encountered numerical issues when using more than two sensors. 
In this method,
each fault is associated a probability density map but it proved impractical to construct one from the available data.

\subsection{DL-FDI in-depth analysis for 10 sensors}
As stated earlier, the FDI is a classification procedure which exploits the discriminative and sparsity properties of the associated dictionary. 
\begin{figure}[!ht]
    \centering
    \includegraphics[width=\columnwidth]{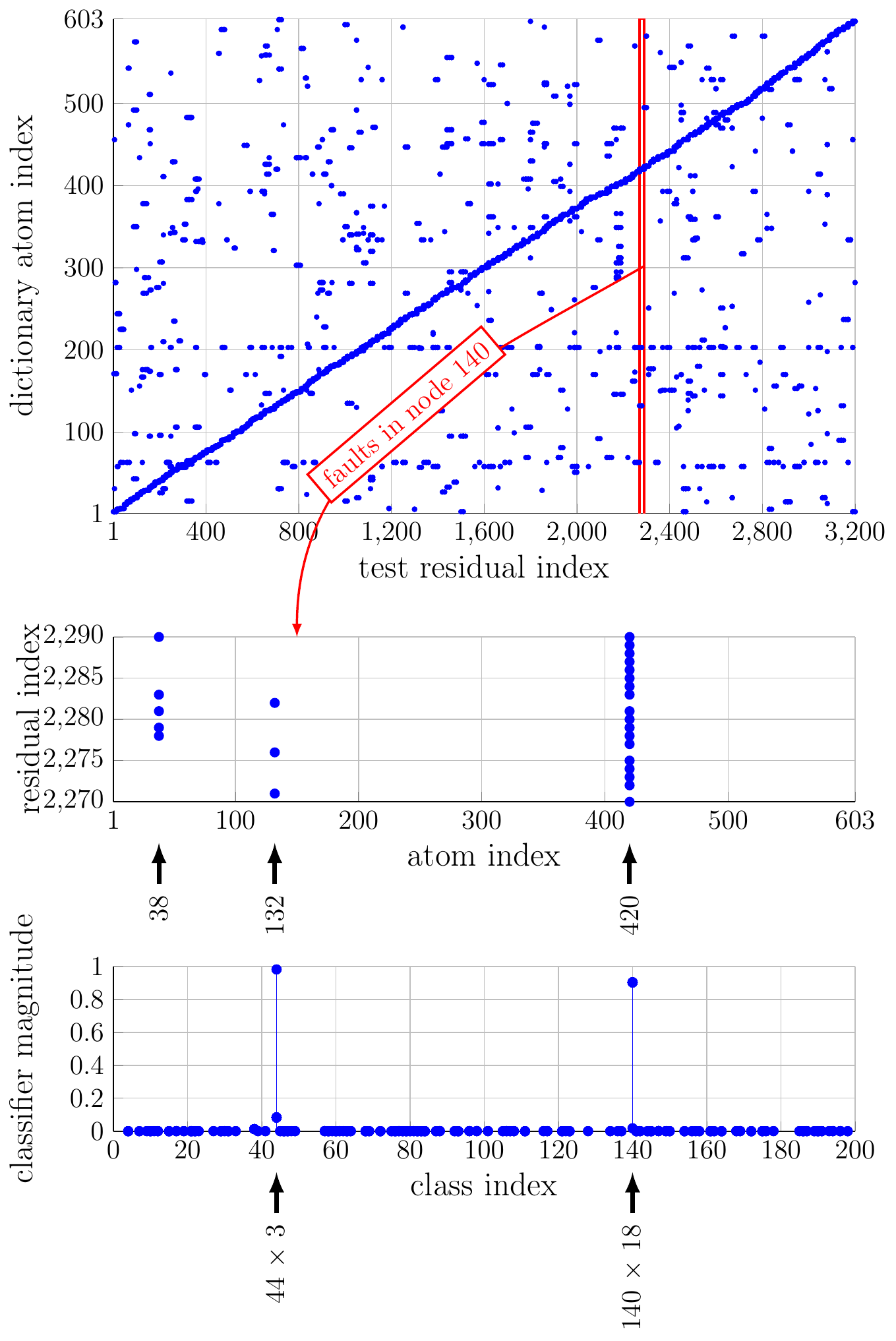}
    \caption{Illustration of dictionary discrimination and sparsity properties (with detail for class 140)}
    \label{fig:dl_vs_x}
\end{figure}

To highlight these properties we illustrate in \figref{fig:dl_vs_x} the active dictionary atoms obtained for the case of $10$ sensors for each of the test residuals considered (a marker at coordinates (i,j) means that in the classification of the i-th residual appears the j-th atom). Note that for a better illustration we re-ordered the test residuals such that the faults appear contiguously.

To better illustrate the approach we take the test residuals corresponding to class 140 (faults affecting node 140), which in \figref{fig:dl_vs_x} correspond to residuals indexed from 2270 to 2290 and show them into the middle inset. We note that a reduced number of atoms ($\{38,132,420\}$) describe the residuals, hence proving the sparsity of the representation. The bottom inset plots the values of the classifier for each of the considered test residuals. We note that the classification returns 3 times class 44 (miss-classification) and 18 times class 140 (correct classification) -- recall that, as per \eqref{class_discrim}, the largest value in $\mathbf W\mathbf x$ indicates the class. For this particular class the success rates are around the average shown in \figref{fig:dl_generic_2_success}, specifically, S1) is $18\cdot 100/21=85.71\%$, S2) and S3) are $21\cdot 100/21=100\%$ since node 44 is the neighbor of node 140.

The diagonal effect in \figref{fig:dl_vs_x} is the result of how matrix $\bm{Q}$ was built.
Recall that the lines in $\bm{Q}$ correspond to the atoms in $\bm{D}$
and its columns to the signals in $\bm{Y}$
and that we set element $\bm{q}_{ij}$ if atom $i$ should represent signal $j$ belonging to class $c$.
This indirectly states that atom $i$ has to represent signals of class $c$.
The signals in \figref{fig:dl_vs_x} were resorted in class-order thus the atom index of the class-specific atoms (dictated by $\bm{Q}$)
also changes every 20 or so residuals resulting in the ascending diagonal aspect.
This is in fact the visual confirmation of the fact that our discrimination strategy worked as residuals might use atoms from the entire dictionary, but they always use at least one from their given class.
    



\section{Conclusions}

We have shown that data-driven approaches can be used successfully for sensor placement and subsequent fault detection and isolation in water networks.
Performing sensor placement through a Gramm-Schmidt-like procedure constrained by the network Laplacian
and then using the resulting sensor data for online dictionary learning
has allowed us to move forward from~\cite{IS17_ifac}
and tackle large networks.
Adaptive learning and classification~\cite{IB19_toddler} 
provides the benefit of a continuous integration of new data into the existing network model,
be it for learning or testing purposes.

The results have shown good accuracy and pointed towards some promising
directions of study such as:
network partitioning into communities,
adapting online dictionary learning to further integrate the network structure (e.g. by enforcing graph smoothness~\cite{Yankelevsky16_dualgraph})
and
providing synergy between the three phases:
placement, learning, FDI
(e.g. allow a flexible placement scheme where the learning iteration is allowed to change the sensor nodes based on current classification results).

\balance
\bibliographystyle{elsarticle-num} 
\bibliography{bib}      
\end{document}